\documentclass[aip,reprint,author-year,nofootinbib,floatfix]{revtex4-2}

\usepackage{physjour_aas_macros}

\usepackage{graphicx}
\usepackage{amsmath,amssymb}             
\usepackage{stmaryrd}					
\usepackage[utf8]{inputenc}
\usepackage[T1]{fontenc}
\usepackage[dvipsnames]{xcolor}
\usepackage{color}
\usepackage[raggedright]{titlesec}
\usepackage[normalem]{ulem}
\usepackage{nicefrac}
\definecolor{darkblue}{rgb}{0.0,0.0,0.4}
\definecolor{darkred}{rgb}{0.7,0.0,0.0}
\definecolor{darkgreen}{rgb}{0.0,0.5,0.0}
\definecolor{C0}{HTML}{1f77b4}
\definecolor{C1}{HTML}{ff7f0e}
\definecolor{C2}{HTML}{2ca02c}
\definecolor{C3}{HTML}{d62728}
\definecolor{C4}{HTML}{9467bd}
\definecolor{C5}{HTML}{8c564b}
\definecolor{C6}{HTML}{e377c2}
\definecolor{C7}{HTML}{7f7f7f}
\definecolor{C8}{HTML}{bcbd22}
\definecolor{C9}{HTML}{17becf}
\usepackage{tikz}
\usetikzlibrary{calc}	
\usepackage{subfigure}

\usepackage{hyperref}
\hypersetup{colorlinks,breaklinks,
            linkcolor=darkblue,urlcolor=darkblue,
            anchorcolor=darkblue,citecolor=RoyalBlue}
\usepackage[capitalise]{cleveref}

\usepackage[mathscr]{euscript}
\usepackage{upgreek}
\DeclareFontFamily{OT1}{pzc}{}
\DeclareFontShape{OT1}{pzc}{m}{it}{<-> s * [1.10] pzcmi7t}{}
\DeclareMathAlphabet{\mathpzc}{OT1}{pzc}{m}{it}

\titleformat{\section}{\selectfont \normalfont\raggedright\sffamily\small\bfseries\uppercase}{\thesection.}{1em}{}{}
\titleformat{\subsection}{\selectfont \normalfont\raggedright\sffamily\small\bfseries}{\thesubsection.}{1em}{}{}
\titleformat{\subsubsection}{\selectfont \normalfont\sffamily\small\bfseries}{\thesubsection.\thesubsubsection}{1em}{}{}
\titleformat{\paragraph}[runin]{\selectfont \normalfont\sffamily\small\bfseries}{\thesubsection.\thesubsubsection.\theparagraph}{1em}{}[.]

\usepackage{tikz}
\usetikzlibrary{shapes,arrows,shadows}
\usepackage[export]{adjustbox}

\usepackage{scalerel}
\usetikzlibrary{svg.path}
\definecolor{orcidlogocol}{HTML}{A6CE39}
\tikzset{
  orcidlogo/.pic={
    \fill[orcidlogocol] svg{M256,128c0,70.7-57.3,128-128,128C57.3,256,0,198.7,0,128C0,57.3,57.3,0,128,0C198.7,0,256,57.3,256,128z};
    \fill[white] svg{M86.3,186.2H70.9V79.1h15.4v48.4V186.2z}
                 svg{M108.9,79.1h41.6c39.6,0,57,28.3,57,53.6c0,27.5-21.5,53.6-56.8,53.6h-41.8V79.1z M124.3,172.4h24.5c34.9,0,42.9-26.5,42.9-39.7c0-21.5-13.7-39.7-43.7-39.7h-23.7V172.4z}
                 svg{M88.7,56.8c0,5.5-4.5,10.1-10.1,10.1c-5.6,0-10.1-4.6-10.1-10.1c0-5.6,4.5-10.1,10.1-10.1C84.2,46.7,88.7,51.3,88.7,56.8z};
  }
}
\newcommand\orcid[1]{\href{https://orcid.org/#1}{\mbox{\scalerel*{
\begin{tikzpicture}[yscale=-1,transform shape]
\pic{orcidlogo};
\end{tikzpicture}
}{|}}}}

\newcommand{\deriv}[2]{\dfrac{\mathrm{d} #1}{\mathrm{d} #2}}

\newcommand{\drm}{\mathrm{d}}

\renewcommand{\i}{\mathrm{i}}

\newcommand{\pylians}{\textsc{pylians}}
\newcommand{\sbmy}{\textsc{Simbelmynë}}
\newcommand{\maptomap}{\textsc{map2map}}
\newcommand{\pytorch}{\textsc{pytorch}}

\begin{document}

\title{COmoving Computer Acceleration (COCA): $N$-body simulations in an emulated frame of reference}



\newcommand{\iap}{CNRS \& Sorbonne Université, UMR 7095, Institut d’Astrophysique de Paris, 98 bis boulevard Arago, F-75014 Paris, France}
\newcommand{\okc}{The Oskar Klein Centre, Department of Physics, Stockholm University, Albanova University Center, SE 106 91 Stockholm, Sweden}
\newcommand{\milan}{Dipartimento di Fisica ``Aldo Pontremoli'', Università degli Studi di Milano, Via Celoria 16, 20133 Milano, Italy}

\author{Deaglan J. Bartlett}
\email{deaglan.bartlett@iap.fr}
\thanks{ORCID: \orcid{0000-0001-9426-7723} \href{https://orcid.org/0000-0001-9426-7723}{0000-0001-9426-7723}; Corresponding author.}
\affiliation{\iap}

\author{Marco Chiarenza}
\affiliation{\iap}
\affiliation{\milan}

\author{Ludvig Doeser}
\affiliation{\okc}

\author{Florent Leclercq}
\email{florent.leclercq@iap.fr}
\homepage{https://www.florent-leclercq.eu/}
\thanks{ORCID: \orcid{0000-0002-9339-1404} \href{https://orcid.org/0000-0002-9339-1404}{0000-0002-9339-1404}; Corresponding author.}
\affiliation{\iap}

\date{\today}

\begin{abstract}
\noindent 
{$N$-body simulations are computationally expensive, so machine-learning-based emulation techniques have emerged as a way to increase their speed.
   Although fast, surrogate models have limited trustworthiness due to potentially substantial emulation errors that current approaches cannot correct for.}
   {To alleviate this problem, we introduce COmoving Computer Acceleration (COCA), a hybrid framework interfacing machine learning with an $N$-body simulator.
   The correct physical equations of motion are solved in an emulated frame of reference, so that any emulation error is corrected by design.
   This approach corresponds to solving for the perturbation of particle trajectories around the machine-learnt solution, which is computationally cheaper than obtaining the full solution, yet is guaranteed to converge to the truth as one increases the number of force evaluations.}
   {Although applicable to any machine learning algorithm and $N$-body simulator, this approach is assessed in the particular case of particle-mesh cosmological simulations in a frame of reference predicted by a convolutional neural network, where the time dependence is encoded as an additional input parameter to the network.}
   {We find that COCA efficiently reduces emulation errors in particle trajectories, requiring far fewer force evaluations than running the corresponding simulation without machine learning. As a consequence, we obtain accurate final density and velocity fields for a reduced computational budget. We demonstrate that this method shows robustness when applied to examples outside the range of the training data. When compared to the direct emulation of the Lagrangian displacement field using the same training resources, COCA's ability to correct emulation errors results in more accurate predictions.
   }
   {Therefore, COCA makes $N$-body simulations cheaper by skipping unnecessary force evaluations, while still solving the correct equations of motion and correcting for emulation errors made by machine learning.}
 
\end{abstract}

\maketitle


\section{Introduction}

$N$-body simulations represent the state-of-the-art numerical method for studying the dynamics of complex systems, including non-linear gravitational structure formation in the Universe \citep{Vogelsberger2020,AnguloHahn2022}.
Such simulations can be incredibly computationally expensive to run \citep[e.g.][]{Potter2017,Heitmann2019,Ishiyama2021,Frontiere2022,Wang2022}; hence, various machine learning (ML)-based approaches have been proposed to either remove the requirement to run physical simulations or to reduce the complexity of the simulator used.

The most straightforward application of ML methods is as surrogate models, which take initial conditions as inputs and emulate various features of the corresponding full $N$-body simulation as outputs.
For example, \citet{LucieSmith2018} were able to predict halo properties from the initial conditions given certain environmental properties.
\citet{Perraudin2019} used Generative Adversarial Networks (GANs) to generate visually plausible three-dimensional cosmological density fields but encountered difficulties in reproducing the correct statistical distribution of physical density fields.
Going further, \citet{Conceicao2024} built an emulator of cosmological density fields based on a combination of dimensionality reduction via principal component analysis and supervised ML.
\citet{He2019} demonstrated that one can replicate the full result of particle-mesh (PM) simulations (i.e., a Lagrangian displacement field) using a deep neural network which takes the Zel'dovich approximation \citep{Zeldovich1970} displacement field as input.
This work was extended to tree-based $N$-body simulations by \citet{deOliveira2020}.
Such emulators can replicate the power spectrum to the percent level up to $k \approx 1 \, h \, {\rm Mpc^{-1}}$.
\citet{Jamieson2023} further extended these works by predicting both the displacement and velocity fields through two separate neural networks and by incorporating the cosmological matter density information through the addition of a ``style'' parameter \citep[][see \cref{sec:architecture}]{Karras2019}.
The resulting emulator can reproduce power spectra and bispectra to within a few percent and achieves a similar level of cross-correlation with the true simulation run with the same initial conditions.
By adding a time variable as an additional style parameter, \citet{Jamieson2024} were able to eliminate the need for two separate networks and produce an emulator capable of predicting $N$-body outputs as a function of redshift.
The speed and differentiable nature of particle-based emulators enable them to be integrated within field-based inference of initial conditions \citep{Doeser2023}, with posterior re-simulations indicating faithful reconstruction of the initial conditions.
Simulations including the effects of massive neutrinos \citep{Giusarma2023} or modified gravity \citep{Saadeh2024} can also be emulated using similar neural network techniques.

Instead of completely bypassing the $N$-body simulation, one can include ML corrections that capture unresolved physics in low-resolution, cheaper simulations.
For example, \citet{Lanzieri2022} introduced an additional effective force to PM simulations to capture unresolved forces between particles.
Their machine-learnt isotropic Fourier filter was extended by \citet{Payot2023} to depend not only on time and wavenumber but also on cosmological parameters.
Super-resolution techniques based on GANs \citep{KodiRamanah2020,Li2021} and U-nets \citep{Zhang2024} have also been proposed, achieving power spectra correct within a few percent, as well as reasonable bispectra, void size functions, and halo mass functions correct to within 10\% for halos down to $\approx 10^{11} \, {\rm M_\odot}$ in mass.
Given the computationally demanding nature of hydrodynamical simulations, \citet{DaiSeljak2021} introduced a light model (with only $\mathcal{O}(10)$ learnable parameters) to transform the output of a dark-matter-only simulation to one that resembles the hydrodynamical simulation run with the same initial conditions.
\citet{Ding2024} also presented a light and interpretable neural network to produce halo catalogues from dark matter density fields.

Accuracy and interpretability are pivotal challenges in the application of machine learning to $N$-body simulations.
Despite the high reported accuracy of the methods reviewed above on various tests (mainly using summary statistics), none of these models can be expected to perfectly recover the truth.
Are ML-accelerated simulation algorithms sufficiently accurate to be used in real-world applications?
Without a ground-truth model to compare against during actual use (since such algorithms are designed to eliminate the need for it), current approaches have limited means of identifying the emulation error and cannot correct for it.
Since typical simulations usually also involve simplifying assumptions and approximations, perfectly accurate ML-based models may not be required for many purposes.
The question that arises is then that of the interpretability of ML, in order to control the approximation made with respect to a physical simulator.
Unfortunately, many ML algorithms, including (deep) neural networks, lack interpretability.
If machines predict something humans do not understand, how can we check (and trust) the results?

In this paper, we contend that addressing the lack of interpretability of ML is not always necessary to use an emulator of an expensive model while maintaining control over the degree of accuracy.
We elucidate this argument by constructing a framework in which emulation of $N$-body simulations is made an ML-safe task by physically rectifying emulation inaccuracies. By ``ML-safe,'' we mean systems that are reliable, robust, and trustworthy \emph{by construction}.
The key idea is to find a mathematically equivalent form of the system's equations of motion, where we solve for the (not necessarily small) perturbation around the approximate solution provided by ML.
From a physical point of view, in $N$-body simulations obeying Newtonian dynamics, this is equivalent to solving the equation of motion in an emulated frame of reference.
Since the ML solution is designed to be approximately correct, computing corrections is numerically easier than evolving the full system, thus requiring fewer evaluations of the forces. 
Through the number and the temporal positions of force evaluations, the user controls the trade-off between speed and accuracy, ranging from fully trusting the ML solution by never correcting particle trajectories to correcting for ML emulation errors at any time step of the simulation.
The system has the theoretical guarantee of asymptotically converging to the physical solution as the number of force evaluations increases.

For gravitational $N$-body simulations of dark matter particles, we introduce the COmoving Computer Acceleration (COCA) approach to running cosmological simulations within an emulated frame of reference.
While traditional emulators aim to translate initial conditions into final particle positions, directly representing the non-linear dark matter distribution, COCA aims to emulate a frame of reference in which to run a physical simulation with lower computational cost.
The approach can be seen as a generalisation and improvement of the idea behind COmoving Lagrangian Acceleration (COLA) \citep{Tassev2013}.
As an illustration, we compare the results of COLA and COCA simulations when forces are evaluated through a particle-mesh (PM) scheme.
The current implementation can be interpreted either as a means of making COLA more computationally efficient (from a simulator's perspective) or as a method for correcting emulators of PM-based simulations (from an emulator builder's perspective).
We find that our ML-enhanced approach requires very few force evaluations to correct for emulation errors: it achieves percent-level accurate power spectra, bispectra, and cross-correlation to a reference simulation, with approximately 8 force evaluations, compared to 20 for COLA.
Importantly, the COCA formalism is not limited to PM codes. Future iterations of this work will extend the framework to other gravity solvers (e.g., P\textsuperscript{3}M or tree-based methods), with the ultimate goal of improving emulators of these more precise codes.

This paper is organised as follows.
In \cref{sec:theory}, we review the COLA approach to $N$-body simulations, extend it to yield COCA, and describe the benefits of COCA in terms of computational efficiency. A more thorough description is provided in \cref{apx:actual equations}.
We introduce our emulator for the frame of reference in \cref{sec:emulator} and describe the training procedure and validation metrics for the COCA simulations. 
In \cref{sec:results}, we present our results: the performance of the emulator, the accuracy of COCA simulations as a function of the number of force evaluations, the generalisation to an example known to be outside the range of the training set, the comparison to a Lagrangian displacement field emulator, and a discussion of the computational performance.
We conclude in \cref{sec:conclusions}, discussing potential future extensions and applications of this study.

\section{Theory}
\label{sec:theory}

For simplicity, some of the equations in this section are abridged. We reintroduce the omitted constants, temporal prefactors, and Hubble expansion in \cref{apx:actual equations}.

\subsection{Review of COLA}

In a cosmological dark matter-only $N$-body code, one wishes to compute the final Eulerian positions of particles $\textbf{x}$, as a function of scale factor $a$, as they interact under gravity. If the initial comoving particle positions are $\textbf{q}$, then the Lagrangian displacement field is given by \citep[e.g.][for a review]{Bernardeau2002}
\begin{equation}
\boldsymbol{\Psi}(\textbf{q},a) \equiv \textbf{x} (a) - \textbf{q}.
\end{equation}
One then must solve the equation of motion which reads schematically
\begin{equation}
\partial_a^2 \boldsymbol{\Psi}(\textbf{q},a) = -\boldsymbol{\nabla} \Phi(\textbf{x},a),
\label{eq:PM_EoM}
\end{equation}
where the gravitational potential $\Phi$ satisfies the Poisson equation sourced by the density contrast field $\delta (\textbf{x}, a)$,
\begin{equation}
\Delta \Phi(\textbf{x},a) = \delta(\textbf{x},a).
\label{eq:Poisson_full_box}
\end{equation}

In the perturbative regime, analytic solutions for $\boldsymbol{\Psi}(\textbf{q},a)$ can be derived, which are known as Lagrangian Perturbation Theory \citep[LPT,][]{Zeldovich1970,Buchert1989,Bouchet1995,Bernardeau2002}. These solutions are valid on large scales but become inaccurate once shell crossing occurs, making the approximation more reliable at early times.
This behaviour is illustrated in \cref{subfig:COLA}, where initially the LPT trajectories and the true trajectories are indistinguishable, but the discrepancy increases over time.

The temporal COmoving Lagrangian Acceleration \citep[][tCOLA or simply COLA in the following]{Tassev2013} algorithm aims to separate the temporal evolution of large and small scales by evolving large scales using analytic LPT results and small scales using a numerical solver. This is accomplished by decomposing the Lagrangian displacement field into two components \citep{TassevZaldarriaga2012}:
\begin{equation}
\boldsymbol{\Psi}(\textbf{q},a) \equiv \boldsymbol{\Psi}_\mathrm{LPT}(\textbf{q},a) + \boldsymbol{\Psi}_\mathrm{res}^\mathrm{COLA}(\textbf{q},a),
\label{eq:displacement_split}
\end{equation}
where
$\boldsymbol{\Psi}_\mathrm{LPT}(\textbf{q},a)$ 
represents the LPT displacement field, and 
$\boldsymbol{\Psi}_\mathrm{res}^\mathrm{COLA}(\textbf{q},a)$ 
denotes the residual displacement of each particle as observed from a frame comoving with an ``LPT observer,'' whose trajectory is defined by $\boldsymbol{\Psi}_\mathrm{LPT}(\textbf{q},a)$. 
Knowing $\boldsymbol{\Psi}_\mathrm{LPT}(\textbf{q},a)$, one does not need to solve for the full trajectory of the particle, but just the residual between the approximation and the truth (the blue arrows in \cref{subfig:COLA}). 

Using \cref{eq:displacement_split}, it is possible to rewrite \cref{eq:PM_EoM} as 
\begin{equation}
\partial_a^2 \boldsymbol{\Psi}_\mathrm{res}^\mathrm{COLA}(\textbf{q},a) = -\boldsymbol{\nabla} \Phi(\textbf{x},a) - \partial_a^2 \boldsymbol{\Psi}_\mathrm{LPT}(\textbf{q},a).
\label{eq:tCOLA_EoM}
\end{equation}
Therefore, one can view LPT as providing a new frame of reference within which we solve the equations of motion. 
The term $\partial_a^2 \boldsymbol{\Psi}_\mathrm{LPT}(\textbf{q},a)$ can be thought of as a fictitious force acting on particles, caused by our use of a non-inertial frame of reference.

Since particles experience lower typical accelerations in the LPT frame compared to the natural cosmological frame, solving the equation of motion numerically becomes a comparatively simpler task, requiring fewer time steps to achieve equivalent accuracy \citep{Tassev2013,Howlett2015,Leclercq2015ST,Koda2016,Izard2016}. In particular, COLA has been demonstrated to always yield correct results at large scales, even with a small number of time steps ($\leq 10$), unlike a basic particle-mesh (PM) code. Given that \cref{eq:tCOLA_EoM} is mathematically equivalent to \cref{eq:PM_EoM}, COLA is asymptotically equivalent to the corresponding standard $N$-body code (e.g. a PM code if forces $-\boldsymbol{\nabla} \left(\Delta^{-1} \delta \right)$ are evaluated via a standard PM technique), in the limit of an infinite number of time steps.

\subsection{COCA formalism}
\label{sec:COCA_formalism}

\begin{figure*}
    \centering
    \subfigure[COLA]{
        \label{subfig:COLA}
        \begin{tikzpicture}[xscale=0.9, yscale=0.9, transform shape, every node/.style={font=\normalsize}]

            \coordinate (start) at (1,1);
            \coordinate (end1) at (9,4);
            \coordinate (end3) at (9,2);

            \coordinate (mid1_1) at (2.6,2.07);
            \coordinate (mid1_2) at (4.2,3.25);
            \coordinate (mid1_3) at (5.8,4.17);
            \coordinate (mid1_4) at (7.4,4.5);

            \coordinate (mid2_1) at (2.7,2.0);
            \coordinate (mid2_2) at (4.6,2.9);
            \coordinate (mid2_3) at (6.2,3.33);
            \coordinate (mid2_4) at (8.0,3.28);

            \coordinate (mid3_1) at (2.75,1.95);
            \coordinate (mid3_2) at (4.7,2.8);
            \coordinate (mid3_3) at (6.2,3.15);
            \coordinate (mid3_4) at (7.9,2.88);

            \coordinate (ml_label) at ($(mid1_4)!0.5!(mid2_4)$);
            \coordinate (res_label) at ($(mid2_4)!0.5!(mid3_4)$);

            \draw[thick, line width=2pt, orange, arrows={-stealth}] (start) to[out=30, in=150] (end1);
            \draw[thick, black, line width=2pt, arrows={-stealth}] (start) to[out=30, in=130] (end3);

            \node at (start) [circle, fill=yellow, inner sep=1.5pt] {};

            \node at (mid1_1) [circle, fill=yellow, inner sep=2pt] {};
            \node at (mid1_2) [circle, fill=yellow, inner sep=2pt] {};
            \node at (mid1_3) [circle, fill=yellow, inner sep=2pt] {};
            \node at (mid1_4) [circle, fill=yellow, inner sep=2pt] {};

            \node at (mid3_1) [circle, fill=yellow, inner sep=2pt] {};
            \node at (mid3_2) [circle, fill=yellow, inner sep=2pt] {};
            \node at (mid3_3) [circle, fill=yellow, inner sep=2pt] {};
            \node at (mid3_4) [circle, fill=yellow, inner sep=2pt] {};

            \draw[blue, ->, line width=1pt] (mid1_1) to (mid3_1);
            \draw[blue, ->, line width=1pt] (mid1_2) to (mid3_2);
            \draw[blue, ->, line width=1pt] (mid1_3) to (mid3_3);
            \draw[blue, ->, line width=1pt] (mid1_4) to (mid3_4);

            \node at (start) [below] {$\mathbf{q}$};
            \node at (end3) [below] {$\mathbf{x}$};
            \node at (end1) [above, orange, yshift=6pt] {LPT};
            \node at (end3) [above, xshift=8pt, yshift=8pt] {$N$-body};
            \node at (res_label) [blue, xshift=6pt, yshift=15pt] {res};

        \end{tikzpicture}
    }
    \subfigure[COCA]{
        \label{subfig:COCA}
        \begin{tikzpicture}[xscale=0.9, yscale=0.9, transform shape, every node/.style={font=\normalsize}]

            \coordinate (start) at (1,1);
            \coordinate (end1) at (9,4);
            \coordinate (end2) at (9,3);
            \coordinate (end3) at (9,2);

            \coordinate (mid1_1) at (2.6,2.07);
            \coordinate (mid1_2) at (4.2,3.25);
            \coordinate (mid1_3) at (5.8,4.17);
            \coordinate (mid1_4) at (7.4,4.5);

            \coordinate (mid2_1) at (2.7,2.0);
            \coordinate (mid2_2) at (4.6,2.9);
            \coordinate (mid2_3) at (6.2,3.33);
            \coordinate (mid2_4) at (8.0,3.28);

            \coordinate (mid3_1) at (2.75,1.95);
            \coordinate (mid3_2) at (4.7,2.8);
            \coordinate (mid3_3) at (6.2,3.15);
            \coordinate (mid3_4) at (7.9,2.88);

            \coordinate (ml_label) at ($(mid1_4)!0.5!(mid2_4)$);
            \coordinate (res_label) at ($(mid2_4)!0.5!(mid3_4)$);

            \draw[thick, line width=2pt, orange, arrows={-stealth}] (start) to[out=30, in=150] (end1);
            \draw[thick, violet, line width=2pt, arrows={-stealth}] (start) to[out=30, in=160] (end2);
            \draw[thick, black, line width=2pt, arrows={-stealth}] (start) to[out=30, in=130] (end3);

            \node at (start) [circle, fill=yellow, inner sep=1.5pt] {};

            \node at (mid1_1) [circle, fill=yellow, inner sep=2pt] {};
            \node at (mid1_2) [circle, fill=yellow, inner sep=2pt] {};
            \node at (mid1_3) [circle, fill=yellow, inner sep=2pt] {};
            \node at (mid1_4) [circle, fill=yellow, inner sep=2pt] {};

            \node at (mid2_1) [circle, fill=yellow, inner sep=2pt] {};
            \node at (mid2_2) [circle, fill=yellow, inner sep=2pt] {};
            \node at (mid2_3) [circle, fill=yellow, inner sep=2pt] {};
            \node at (mid2_4) [circle, fill=yellow, inner sep=2pt] {};

            \node at (mid3_1) [circle, fill=yellow, inner sep=2pt] {};
            \node at (mid3_2) [circle, fill=yellow, inner sep=2pt] {};
            \node at (mid3_3) [circle, fill=yellow, inner sep=2pt] {};
            \node at (mid3_4) [circle, fill=yellow, inner sep=2pt] {};

            \draw[darkgreen, ->, line width=1pt] (mid1_1) to (mid2_1);
            \draw[blue, ->, line width=1pt] (mid2_1) to (mid3_1);
            \draw[darkgreen, ->, line width=1pt] (mid1_2) to (mid2_2);
            \draw[blue, ->, line width=1pt] (mid2_2) to (mid3_2);
            \draw[darkgreen, ->, line width=1pt] (mid1_3) to (mid2_3);
            \draw[blue, ->, line width=1pt] (mid2_3) to (mid3_3);
            \draw[darkgreen, ->, line width=1pt] (mid1_4) to (mid2_4);
            \draw[blue, ->, line width=1pt] (mid2_4) to (mid3_4);

            \node at (start) [below] {$\mathbf{q}$};
            \node at (end3) [below] {$\mathbf{x}$};
            \node at (end1) [above, orange, yshift=6pt] {LPT};
            \node at (end2) [above, violet, yshift=6pt] {emulated};
            \node at (end3) [above, xshift=8pt, yshift=8pt] {$N$-body};
            \node at (res_label) [blue, xshift=8pt, yshift=-1pt] {res};
            \node at (ml_label) [darkgreen, xshift=9pt, yshift=1pt] {ML};

        \end{tikzpicture}
    }
    \caption{Schematic illustration of the (a) COLA and (b) COCA formalism for cosmological simulations. In COLA, one solves the equations of motion in the frame of reference given by LPT, so one computes the residual (``res'') between the LPT trajectory and the true position $\textbf{x}$, of particles. In COCA, one emulates a frame of reference closer to the true trajectory by adding a ML contribution to LPT, so one solves for the (smaller) residuals between $\textbf{x}$ and the emulated frame.}
    \label{fig:coca_sketch}
\end{figure*}
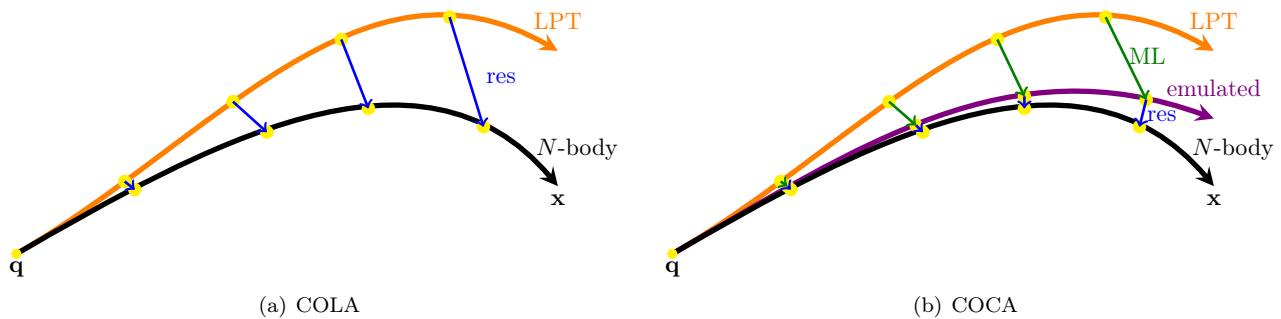

While the COLA formalism has proven effective in solving the equation of motion within the frame of reference defined by LPT, there is no requirement to adhere to LPT or any other analytic approximation in the decomposition given by \cref{eq:displacement_split}.
According to the principle of Galilean invariance, the equation of motion can be solved in \textit{any} frame of reference, provided appropriate fictitious forces are introduced for non-inertial frames.
Considering that the simplest scenario is the one where no motion occurs, we aim at finding a frame of reference where particles are nearly stationary. In such a frame of reference, solving the equation of motion numerically to reach a given level of accuracy becomes an easier problem than in COLA. This is the key insight that underpins the formalism proposed in this paper. We dub this approach COmoving Computer Acceleration (COCA).

We propose utilising a ML algorithm, such as a neural network, as an emulator to learn and predict the trajectories of particles in $N$-body simulations. Since the LPT frame already provides a good approximation of the trajectories, particularly on large scales, we opt to learn displacements relative to the LPT frame. Therefore, the emulator outputs a displacement field $\boldsymbol{\Psi}_\mathrm{ML}(\textbf{q},a)$ that approximates $\boldsymbol{\Psi}(\textbf{q},a) - \boldsymbol{\Psi}_\mathrm{LPT}(\textbf{q},a)$.

Rather than directly employing the emulator as a surrogate for simulation results, we use the frame of reference corresponding to the emulated trajectories in order to run a simulation. Following the same spirit as COLA, we split the Lagrangian displacement field into three contributions,
\begin{equation}
\boldsymbol{\Psi}(\textbf{q},a) \equiv \boldsymbol{\Psi}_\mathrm{LPT}(\textbf{q},a) + \boldsymbol{\Psi}_\mathrm{ML}(\textbf{q},a) + \boldsymbol{\Psi}_\mathrm{res}^\mathrm{COCA}(\textbf{q},a),
\label{eq:displacement_split_COCA}
\end{equation}
where $\boldsymbol{\Psi}_\mathrm{ML}(\textbf{q},a)$ is the ML contribution to the Lagrangian displacement field, and the residual displacement $\boldsymbol{\Psi}_\mathrm{res}^\mathrm{COCA}(\textbf{q},a)$ represents the emulation error. Different contributions are shown schematically in \cref{subfig:COCA}.

Reframing \cref{eq:PM_EoM} using \cref{eq:displacement_split_COCA}, the equation of motion for COCA contains an extra fictitious force with respect to COLA:
\begin{equation}
\partial_a^2 \boldsymbol{\Psi}_\mathrm{res}^\mathrm{COCA}(\textbf{q},a) = -\boldsymbol{\nabla} \Phi(\textbf{x},a) - \partial_a^2 \boldsymbol{\Psi}_\mathrm{LPT}(\textbf{q},a) 
- \partial_a^2 \boldsymbol{\Psi}_\mathrm{ML}(\textbf{q},a).
\label{eq:COCA_EoM}
\end{equation}

In COCA, the predicted displacement $\boldsymbol{\Psi}_\mathrm{LPT}(\textbf{q},a) + \boldsymbol{\Psi}_\mathrm{ML}(\textbf{q},a)$ approximates the optimal frame of reference in which to solve the simulation (the one where all particles are at rest). Ideally, in case of perfect emulation, solving the equation of motion would result in no trajectory adjustment ($\boldsymbol{\Psi}_\mathrm{res}^\mathrm{COCA}(\textbf{q},a) =0$ for any $a$). Otherwise, numerically solving \cref{eq:COCA_EoM} corrects the trajectories of particles to produce a more accurate solution.

We describe in more detail the COCA formalism in \cref{apx:actual equations}. Notably, while above we described the framework in terms of an emulated displacement field $\boldsymbol{\Psi}_\mathrm{ML}(\textbf{q},a)$, we show that we can equivalently define the new frame of reference by the momentum $\textbf{p} \equiv \mathrm{d}\textbf{x}/\mathrm{d}a$ of particles, so that
\begin{equation}
\textbf{p}(a) \equiv \textbf{p}_\mathrm{LPT}(a) + \textbf{p}_\mathrm{ML}(a) + \textbf{p}_\mathrm{res}(a),
\end{equation}
where $\textbf{p}_\mathrm{LPT}(a)$ and $\textbf{p}_\mathrm{ML}(a)$ denote momenta predicted by LPT and ML, respectively, and the residual momentum $\textbf{p}_\mathrm{res}(a)$ is determined by solving the equations of motion.

\subsection{Reducing the number of force evaluations}

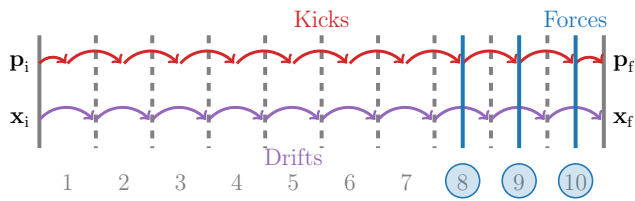
\begin{figure}
    \resizebox{\linewidth}{!}{
    \begin{tikzpicture}[every node/.style={font=\large}]

        \tikzset{
            drift/.style={->, draw=C4, thick, line width=1.5pt},
            kick/.style={->, draw=C3, thick, line width=1.5pt},
            force/.style={-, draw=C0, thick, line width=2pt},
            textstyle/.style={font=\large},
            dashedline/.style={dash pattern=on 4pt off 4pt, draw=gray, line width=2pt},
            solidline/.style={draw=gray, line width=2pt}
        }

        \foreach \i in {1,...,9} {
            \draw[dashedline] (\i + 0.5, 1.5) -- (\i + 0.5, -0.5);
        }
        \draw[solidline] (0.5, 1.5) -- (0.5, -0.5);
        \draw[solidline] (10.5, 1.5) -- (10.5, -0.5);

        \foreach \i in {1,...,10} {
            \draw[drift] (\i-0.5, 0) arc[start angle=140, end angle=40, radius=0.62];
            \ifnum\i<10
                \draw[kick] (\i, 1) arc[start angle=140, end angle=40, radius=0.62];
            \fi
        }
        \draw[kick] (0.5, 1) arc[start angle=140, end angle=40, radius=0.31];
        \draw[kick] (10.0, 1) arc[start angle=140, end angle=40, radius=0.31];

        \node[left] at (0.45, 1) {$\textbf{p}_{\rm i}$};
        \node[left] at (0.45, 0) {$\textbf{x}_{\rm i}$};
        \node[right] at (10.55, 1) {$\textbf{p}_{\rm f}$};
        \node[right] at (10.55, 0) {$\textbf{x}_{\rm f}$};

        \foreach \i in {8, 9, 10} {
            \draw[force] (\i, 1.5) -- (\i, -0.5);
            \draw[C0, thick] (\i, -1.08) circle (0.30);
        }

        \foreach \i in {1,...,10} {
            \node[below, gray] at (\i, -0.85) {\i};
        }
        
        \node[textstyle, C4] at (5.0, -0.65) {Drifts};
        \node[textstyle, C3] at (5.5, 1.8) {Kicks};
        \node[textstyle, C0] at (10, 1.8) {Forces};

    \end{tikzpicture}
    }
    \caption{Schematic illustration of the kick-drift-kick integration scheme employed in this study. The initial positions $\textbf{x}_\mathrm{i}$ and momenta $\textbf{p}_\mathrm{i}$ are evolved to their final values $\textbf{x}_\mathrm{f}$ and $\textbf{p}_\mathrm{f}$, with updates to these quantities occurring at different times. Unlike traditional kick-drift-kick integration, we choose not to evaluate forces that appear in the equations of motion at all time steps, but only at a subset (steps 8, 9 and 10 in this example). At all other kick steps, the momenta are updated according to the emulated frame of reference only.
    }
    \label{fig:leapfrog_sketch}
\end{figure}

To integrate the equations of motion in the new frame of reference, we utilise a symplectic ``kick-drift-kick'' (leapfrog) algorithm \citep[e.g.][]{Birdsall1985}. 
With this method, the positions $\textbf{x}$, and momenta $\textbf{p}$, of the particles are updated at different times, typically with one momentum update between every two position updates.
A schematic illustration of the technique is given in \cref{fig:leapfrog_sketch}, with the full details provided in \cref{apx:actual equations}.

At each momentum update (``kick'') we face two choices. One can assume that the emulated frame of reference is sufficiently accurate and thus update the particle momenta by simply evaluating the emulator, corresponding to following the ``emulated'' (purple) trajectory in \cref{subfig:COCA} (equivalent to assuming that $\textbf{g}_\delta(t^\mathrm{D}) = \textbf{0}$ in equation \eqref{eq:K_generic}, using the notations of \cref{apx:actual equations}).
Alternatively, one may deem the emulation error significant and opt to correct the trajectory, aiming to bring the particles back to the ``$N$-body'' (black) trajectory in \cref{subfig:COCA}.
This correction involves evaluating gravitational forces between particles\footnote{In a PM scheme, which we employ in this paper, forces are computed by deriving the density field from particle positions through cloud-in-cell binning, solving the Poisson equation in Fourier space to obtain the gravitational potential, and then finite differencing the potential in configuration space to get the forces.} ($\textbf{g}_\delta(t^\mathrm{D})$ in \cref{apx:actual equations}) and using the complete form of the kick operator.
Correcting trajectories is more computationally expensive than simply following the emulated trajectories, so the number of force evaluations $n_\mathrm{f}$, should be as small as possible, but large enough to correct for emulation errors.
During time steps without force evaluations, particles move according to trajectories defined by their respective frames of reference ($\boldsymbol{\Psi}_\mathrm{LPT}$ for COLA and $\boldsymbol{\Psi}_\mathrm{LPT}+\boldsymbol{\Psi}_\mathrm{ML}$ for COCA). Hence, $n_\mathrm{f}=0$ corresponds to the LPT solution in COLA simulations and a purely emulated one in COCA simulations.

In COCA, the ability to reduce the number of force evaluations introduces an additional degree of freedom compared to PM/COLA simulations, where forces are evaluated at every time step. Force evaluations can in principle be placed at any of the time steps, however we find that concentrating all evaluations towards the end of the simulation, when structure formation is non-linear, typically yields the most accurate results.
Up until the first force evaluation, the COCA framework consists in predicting particle positions $\textbf{x}$ and momenta $\textbf{p}$ at specific times, functioning in a similar way as more traditional emulators \citep[e.g.][]{Jamieson2023}.
In \cref{fig:leapfrog_sketch} we show an example of a kick-drift-kick scheme with ten time steps, with three force evaluations at time steps 8, 9 and 10. At all other time steps, momentum updates (kicks) rely solely on the chosen frame of reference.

\section{Emulation}
\label{sec:emulator}

We remind the reader that the field to be emulated, $\textbf{p}_\mathrm{ML}(\textbf{q},a)$, is the residual momentum field with respect to the LPT momentum field, namely
\begin{equation}
\textbf{p}(\textbf{q},a) - \textbf{p}_\mathrm{LPT}(\textbf{q},a),
\end{equation}
at any value of $a$ corresponding to a kick time step (see \cref{apx:actual equations}).

\subsection{Training data}
\label{sec:training_data}

For the application of COCA described in this work, we chose to emulate the frame of reference in a cubic box of length $128 \, h^{-1} \, {\rm Mpc}$ with $N^3 = 64^3$ dark matter particles, resulting in a final density field at $a=1$ on a grid with a resolution of $\Delta x = 2 \, h^{-1} \, {\rm Mpc}$. This resolution is approximately the same as that used by \citet{Jamieson2023}.\footnote{
The number of force evaluations needed in COCA to achieve a given accuracy likely depends on the accuracy of the frame of reference emulator and, therefore, on the resolution. We leave the investigation of such effects to future work.
}
Since the focus of this paper is the time evolution of the fields, we adopt fixed cosmological parameters equal to the best-fit values (TT,TE,EE+lowE+lensing+BAO) from Planck 2018 \citep{Planck2018VI}: $\Omega_{\rm b} = 0.0490$, $\Omega_{\rm m} = 0.3111$, $h = 0.6766$, $\tau = 0.0561$, $n_{\rm s} = 0.9665$, and $\sigma_8 = 0.8102$. We assume a flat Universe and a non-evolving equation of state for dark energy.

Although the COCA formalism can be applied to any method of computing the forces between particles (PM, P$^3$M, tree-based, etc.), for this paper, we chose to work with a PM force solver, utilising a modified version of the publicly available \sbmy{} code\footnote{\url{https://simbelmyne.florent-leclercq.eu/}} \citep{Leclercq2015ST, Leclercq2020sCOLA}.
For our simulations, we generated initial conditions at a scale factor $a = 0.05$ using second-order LPT and solved the equations of motion using COLA with 20 time steps equally spaced in $a$ and a PM grid of size $64^3$ \citep[see][for investigations on the effect of these choices in COLA simulations]{Izard2016,Koda2016}.
Although we have verified that this initial scale factor and number of time steps are appropriate to give converged results for all $k \leq 1 \, h \, {\rm Mpc}^{-1}$, the ``reference'' against which we compare in testing refers to a COLA simulation with the same setup, except with 100 time steps equally spaced in $a$.
At each time step of the simulations, we output the difference between the computed momentum of the particles $\textbf{p}$ and the LPT momentum $\textbf{p}_{\rm LPT}$, which is the quantity we must emulate.

We produce 100 simulations for training, 50 for validation, and a further 50 for testing.
This is a sufficiently small number of training simulations that re-training with a different resolution or specifications does not require significant computational resources.
While one could potentially achieve higher accuracy for the emulator with more training simulations, the aim of this paper is primarily to demonstrate how to correct for emulation errors rather than to produce the optimal emulator. Therefore, we find 100 training simulations to be sufficient for our purposes.
For each simulation, we use all 20 output snapshots, resulting in a total of 2000 fields for training. In addition, we use 1000 fields for validation and 1000 for testing.

\subsection{Scaling of momenta}

\begin{figure*}
    \centering
    \includegraphics[width=\textwidth]{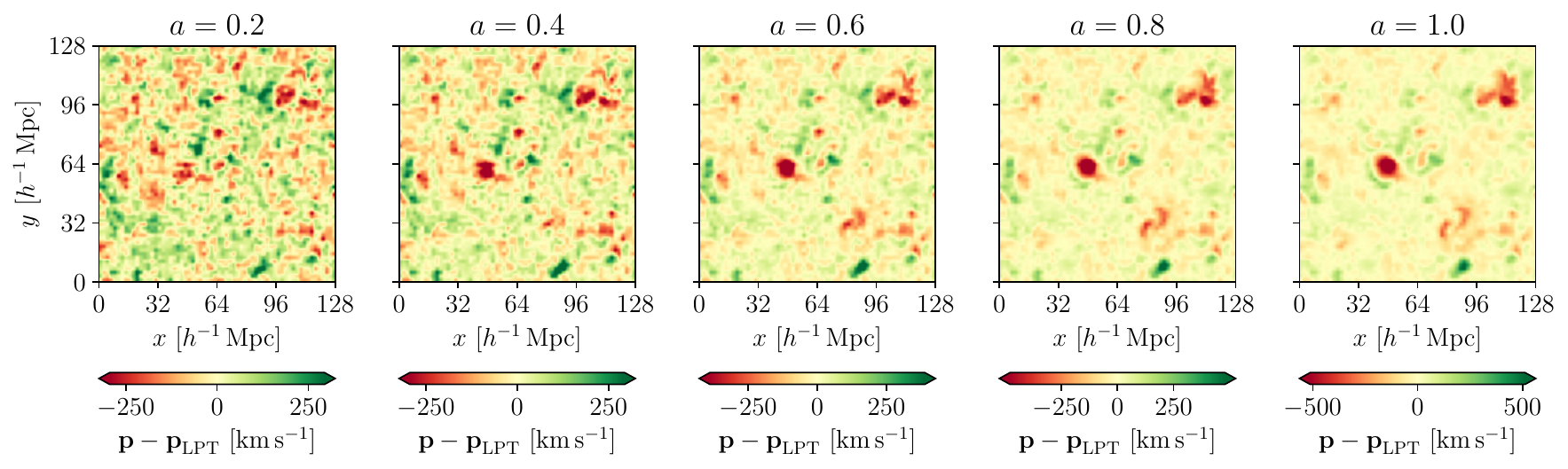}
    \caption{
    Slice of the difference between the true momenta of particles $\textbf{p}$ and the LPT prediction $\textbf{p}_{\rm LPT}$, for a test simulation, as a function of scale factor $a$. We plot the component orthogonal to the chosen slice. At late times, the spatial structure of the field $\textbf{p} - \textbf{p}_{\rm LPT}$ remains relatively constant, with most of the time dependence being a simple multiplicative scaling.
}
    \label{fig:time_dependent_pres}
\end{figure*}

In \cref{fig:time_dependent_pres}, we plot one component of the field $\textbf{p} - \textbf{p}_{\rm LPT}$ as a function of scale factor, for a slice of one of our test simulations.
From visual inspection, we find that the large-scale spatial structure of the field to be emulated does not change significantly as a function of time, particularly at late times, but its magnitude does.
We therefore choose to rescale the momenta to be emulated by defining
\begin{equation}
    \label{eq:momentum_scaling_definition}
    \textbf{p}_{\rm ML} \left( \textbf{q}, a \right)
    \equiv
    D(a) \mathcal{H}(a) \varpi (a) \tilde{\textbf{p}}_{\rm ML} \left( \textbf{q}, a \right),
\end{equation}
where $\tilde{\textbf{p}}_{\rm ML}$ is defined to have a standard deviation of unity, $D(a)$ is the linear growth factor, $\mathcal{H}(a)$ is the conformal Hubble parameter in units of $h$, and $\varpi(a)$ is a time-dependent function which we wish to approximate.
Our emulator is designed to directly predict $\tilde{\textbf{p}}_{\rm ML}$, and thus this scaling has the benefit of standardising the output, since $\tilde{\textbf{p}}_{\rm ML}$ has zero mean and standard deviation unity.

\begin{figure}
    \centering
    \includegraphics[width=\columnwidth]{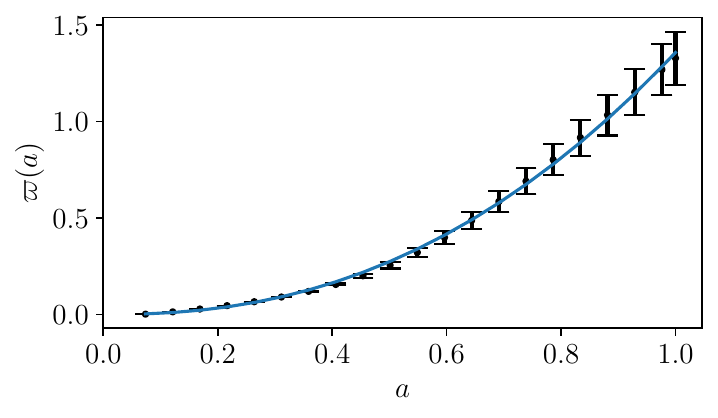}
    \caption{Scaling of the residual momentum as a function of scale factor, as defined in \cref{eq:momentum_scaling_definition}. The points and error bars represent the mean and standard deviation, respectively, across the 100 training simulations. The curve represents the best fit, as given by \cref{eq:momentum_scaling_fit}.    
    }
    \label{fig:momentum_scaling}
\end{figure}

To find an approximation for $\varpi(a)$, we compute the standard deviation of the 2000 training $\textbf{p} - \textbf{p}_{\rm LPT}$ fields and fit these as a function of $a$ using the ESR \citep{Bartlett2023ESR} symbolic regression code.
We use a mean squared error loss function and allow functions to be comprised of addition, multiplication, subtraction, division, the power operator, as well as free constants, $\theta$, and the scale factor $a$.
Upon inspecting the fitted equations, we find that a power law provides a sufficiently simple yet accurate approximation for our purposes:
\begin{equation}
	\label{eq:momentum_scaling_fit}
	\varpi (a) \approx  \left( \theta_0 a \right)^{\theta_1},
\end{equation}
with parameters $\theta_0 = 1.1415174$ and $\theta_1 = 2.3103984$, which yields a root mean squared error of $1.5 \times 10^{-3}$.
We compare this fit to the training data in \cref{fig:momentum_scaling}, from which we see that it accurately reproduces $\varpi(a)$ at all scale factors. Note that any error in this fit can be compensated for by the emulator, so a perfect fit is not required.

\subsection{Neural network architecture}
\label{sec:architecture}

To emulate $\tilde{\textbf{p}}_{\rm ML}$, we utilise a U-net/V-net architecture \citep{Ronneberger2015,Milletari2016}, with a similar implementation to \citet{deOliveira2020,Jamieson2023,Jamieson2024}.
Our model consists of three resolution levels connected in a ``V'' shape, using two downsampling (by stride-2 $2^3$ convolutions) and two upsampling (by stride-$\nicefrac{1}{2}$ $2^3$ convolutions) layers. At each level, we apply a $3^3$ convolution and, as in a V-Net, we apply a $1^3$ convolution as a residual connection \citep[ResNet,][]{He2016} within each block. 
A batch normalisation is applied after each convolution, which is followed by a leaky ReLU activation function with a negative slope of 0.01. Each layer has 64 channels, except the input (1), output (3), and those after concatenations (128).

For every convolutional layer, we introduce a ``style'' parameter\footnote{
A style parameter in a neural network is an additional input at each layer that encodes dependence on an important feature. In our case, the scale factor $a$ encodes the dependence on clustering at various cosmological times. For more details, we refer the interested reader to Eq. (1)--(3) of \citet{Karras2019}.
} 
(borrowing the nomenclature from StyleGAN2, \citealp{Karras2019}), where each convolutional kernel is multiplied by an array of the same dimension as the layer's input and with values equal to the style parameter.
Since we are producing a time-dependent emulator, we use the scale factor $a$ as a style parameter \citep[similar to][who also include a redshift-dependent style parameter]{Jamieson2024}.
Our network is implemented and trained using a modified version of the \maptomap\footnote{\url{https://github.com/eelregit/map2map/}} package and \pytorch{} \citep{Paszke2019}.

The input to the emulator is the redshift-zero linear density field (at $a=1$). This contrasts with \citet{He2019}, \citet{deOliveira2020}, \citet{Jamieson2023} and \citet{Jamieson2024}, who use the three displacement fields predicted by first-order LPT as input. Given that the latter is computed deterministically from the former, both fields contain the same amount of information. However, the linear density field requires three times less memory to store, making our approach more memory efficient.

Compared to previous $N$-body emulators, our architecture is most similar to that of \citet{Jamieson2023}, although we use only $N^3=64^3$ voxels in the input field, whereas \citet{Jamieson2023} employ $N^3=128^3$ voxels with a similar voxel size.
This reduction in voxel count decreases memory requirements by an additional factor of 8 compared to \citet{Jamieson2023}.
Furthermore, we require less padding of the input field than \citet{Jamieson2023}: our approach employs periodic padding of 24 voxels, compared to 48 in their implementation.
The smaller input size (resulting from using a scalar field rather than a three-vector field) necessitates one fewer resolution layer in our neural network architecture, thus reducing the number of trainable parameters in our model to $2.4\times10^6$, compared to $3.4\times10^6$ in \citet{Jamieson2023}.
\citet{He2019}, who emulate the displacement field of particle-mesh simulations, use $N^3=32^3$ voxels with the same box size as in our work, resulting in a voxel size that is twice as large. Despite their lower resolution, their architecture employs $8.4\times10^6$ parameters.
It is worth noting that while our model has fewer parameters than \citet{Jamieson2023}, unlike their emulator, our network does not currently encode dependence on cosmological parameters. A similar number of parameters may be necessary once this feature is incorporated.

Regarding the dependence of the emulation on cosmology, we expect the sensitivity of $\textbf{p} - \textbf{p}_\mathrm{LPT}$ to cosmological parameters to be relatively small, since long-range features should be captured in $\textbf{p}_{\rm LPT}$.
Moreover, we choose to use the linear density field as input instead of the white noise field from which it is produced. This way, our emulator of $\textbf{p} - \textbf{p}_\mathrm{LPT}$ only depends on $\Omega_{\rm m}$, as the equations of motion depend solely on this parameter.
The dependence on all other cosmological parameters is contained in the linear power spectrum, which is used to transform the white noise field into the linear density field. 
Thus, adding only $\Omega_{\rm m}$ as a second style parameter to the network would be sufficient to capture the dependence of the framework on cosmological parameters. For simplicity, we fix $\Omega_{\rm m}$ and save this extension for future work.
Omitting $\Omega_{\rm m}$ as a second style parameter also enables us to test the robustness of the COCA framework in the case of cosmological parameter misspecification, and hence check for ML-safety.
We discuss this aspect in \cref{sec:ml-safety}.

To summarise, our architecture is similar to that of \citet{Jamieson2023}, with three main differences:
(i) we use a single channel (linear density) input rather than three channels (LPT displacements or velocities);
(ii) we have three resolution levels instead of four (since we work with $N^3=64^3$ grids as opposed to $N^3=128^3$); and
(iii) we include $a$ as a style parameter 
\citep[as in][]{Jamieson2024} 
and fix $\Omega_{\rm m}$.

While it is possible to discuss similarities and differences in the neural network architectures, we cannot directly compare the output of our emulator with that of \citet{He2019,Jamieson2023} or \citet{Jamieson2024}. This is because these works emulate the Lagrangian displacement field $\boldsymbol\Psi$, whereas our emulator predicts the residual momentum $\tilde{\textbf{p}}_{\rm ML}$, the quantity required by the COCA framework.
We note that \citet{Jamieson2023,Jamieson2024} employ a more accurate gravity solver in their training simulations compared to the PM model used by \citet{He2019} and in this work. However, this difference does not affect the respective neural network architectures.

\subsection{Training}
\label{sec:training}

As our loss function, we choose
\begin{equation}
    {\rm Loss} \equiv \log L_1 + \log L_2,
\end{equation}
where
\begin{equation}
    L_n \equiv \sum_{\textbf{q}} \sum_i
        \left\lbrace
      \left[\left( \textbf{p}_{\rm LPT} + \textbf{p}_{\rm ML} \right)_i\right]^n - 
      \left[\left( \textbf{p}_{\rm true} \right)_i \right]^n
      \right\rbrace^2,
\end{equation}
and the sum runs over the Lagrangian coordinates of the particles, $\textbf{q}$, and the three Cartesian components, $i \in { x, y, z}$.

This functional form is partially inspired by \citet{Jamieson2023}.
The $L_1$ term matches $\textbf{p}_\mathrm{ML}$ to the residual momenta $\textbf{p}_\mathrm{true} - \textbf{p}_\mathrm{LPT}$, whereas the $L_2$ term ensures that the full momentum field (including the LPT contribution) matches $\textbf{p}_\mathrm{true}$.
\citet{Jamieson2023} found that terms similar to $L_2$ are required to correctly predict redshift-space distortions. We leave the investigation of redshift-space distortions in COCA for future work.
Both terms of our loss function use the mean square error between the fields in Lagrangian coordinates. Unlike \citet{Jamieson2023}, we do not include any term in Eulerian coordinates. Given the computational and memory requirements to use the displacement fields in Eulerian coordinates, and the good performance already achieved with our choice, we decided to omit such additional terms.

We use the Adam optimiser \citep{Kingma2014} with decoupled weight decay (AdamW) \citep{Loshchilov2017}, an initial learning rate of $1.5\times10^{-4}$, a weight decay coefficient of $8\times10^{-3}$, and parameters $\beta_1=0.85$, $\beta_2 = 0.994$, and $\epsilon = 3\times10^{-9}$.
The learning rate is reduced on a plateau by a factor of 0.35 when the loss does not improve by more than $10^{-3}$ over 50 epochs. After a change in learning rate, we apply a cooldown of 30 epochs before the scheduler resumes normal operation. We use a batch size of 5 and train on a single V100 GPU, which has 32~GB of RAM.
The entire time for generating the training, validation, and test simulations (for which we use 40 Intel Xeon Gold 6230 cores) and training was 120 hours, corresponding to 277 epochs, by which time the training and validation losses had plateaued.

\subsection{Validation metrics}
\label{sec:metrics}

To quantitatively determine the accuracy of the COCA simulations, we compute the dark matter density field $\delta$ using a cloud-in-cell estimator \citep{Hockney1981} and the velocity field $\textbf{v}$ using a simplex-in-cell estimator \citep{Hahn2015,Leclercq2017}. To work with a scalar field rather than a vector field, we compute the divergence of the velocity field in Fourier space, $\nabla \cdot \textbf{v}$.\footnote{
The velocity potential is usually of greater physical interest than the divergence of the velocity field. However, in Fourier space, they are related by a factor of $1/k^2$, and since we only compare the ratio of auto and cross spectra at a given $k$, all quantities shown will be identical for both. Thus, we compute only the divergence.
}

For both fields $\varphi \in \{ \delta, \nabla \cdot \textbf{v} \}$, we compute the (auto) power spectrum $P_\varphi (k)$ defined by
\begin{equation}
    \langle \varphi \left( \textbf{k} \right) \varphi \left(  \textbf{k}^\prime \right) \rangle 
    \equiv \left(2\pi\right)^3 \updelta_{\rm D} \left(  \textbf{k} +  \textbf{k}^\prime \right) P_\varphi (k),
\end{equation}
where $\updelta_{\rm D}$ is a Dirac delta distribution. 
For all simulations, we compute the ratio of power spectra between the simulation of interest and the reference.
We also compute the cross spectrum $P_{\varphi_a \varphi_b} (k)$ between the test simulation and the reference simulation, defined by
\begin{equation}
    \langle \varphi_a \left( \textbf{k} \right) \varphi_b \left(  \textbf{k}^\prime \right) \rangle 
    \equiv \left(2\pi\right)^3 \updelta_{\rm D} \left(  \textbf{k} +  \textbf{k}^\prime \right) P_{\varphi_a \varphi_b} (k).
\end{equation}
Thus, we obtain the cross-correlation coefficient
\begin{equation}
    r_{\varphi_a \varphi_b} \left( k \right)
    = \frac{P_{\varphi_a \varphi_b} (k)}{\sqrt{P_{\varphi_a} (k) P_{\varphi_b} (k)}} .
\end{equation}
One can interpret $1-r^2$ as the fraction of the variance in the prediction that is not explained by the reference.
In schemes such as \textsc{carpool} \citep{Chartier2021}, where one combines exact and approximate simulations, $1-r^2$ is proportional to the required number of simulations. Hence, improving $r^2$ can dramatically reduce the required computational resources. 
Just comparing $r$ can hide the importance of improving the cross-correlation: for example, improving $r$ from 0.9 to 0.99---a change of 0.09---corresponds to explaining an additional 17\% of the variance at that scale. 
For these reasons, in all figures, we plot $r^2$ rather than $r$, since it is more meaningful.
All two-point statistics are computed using \sbmy{}.

To assess higher-order statistics, we also compute the bispectrum $B (k_1, k_2, k_3)$ of the density field, defined by
\begin{equation}
    \langle \delta \left( \textbf{k}_1 \right)\delta \left(  \textbf{k}_2 \right) \delta\left(  \textbf{k}_3 \right) \rangle 
    \equiv \left(2\pi\right)^3 \updelta_{\rm D} \left( \sum_{i=1}^3 \textbf{k}_i  \right) B (k_1, k_2, k_3),
\end{equation}
and, to factor out dependence on scale and cosmological parameters, the reduced bispectrum,
\begin{equation}
    \label{eq:reduced_bispectrum}
    Q (k_1, k_2, k_3)
    \equiv \frac{B (k_1, k_2, k_3)}{P_1 P_2 + P_2 P_3 + P_3 P_1},
\end{equation}
with $P_i \equiv P_\delta(k_i)$ for $i \in \left\lbrace 1,2,3 \right\rbrace$.
We consider two different configurations in this work, which are designed to be approximately the same as those used in \citet{Jamieson2023,Doeser2023}. First, we consider a ``squeezed'' bispectrum, consisting of an isosceles triangle configuration with one small wavenumber, $k_{\rm \ell} = 9.8 \times 10^{-2} \, h{\rm \, Mpc}^{-1}$, and two sides of equal but varying size, $k_1=k_2=k_{\rm s}$. For our second configuration, we fix two of the wavenumbers, $k_1 = 0.1 \, h{\rm \, Mpc}^{-1}$ and $k_2 = 1.0 \, h{\rm \, Mpc}^{-1}$, and vary the angle $\theta$ between them.
All bispectrum calculations are performed using \pylians{} \citep{Pylians}.

\section{Results}
\label{sec:results}

\subsection{Emulator performance}

\begin{figure*}
    \centering
    \includegraphics[width=\textwidth]{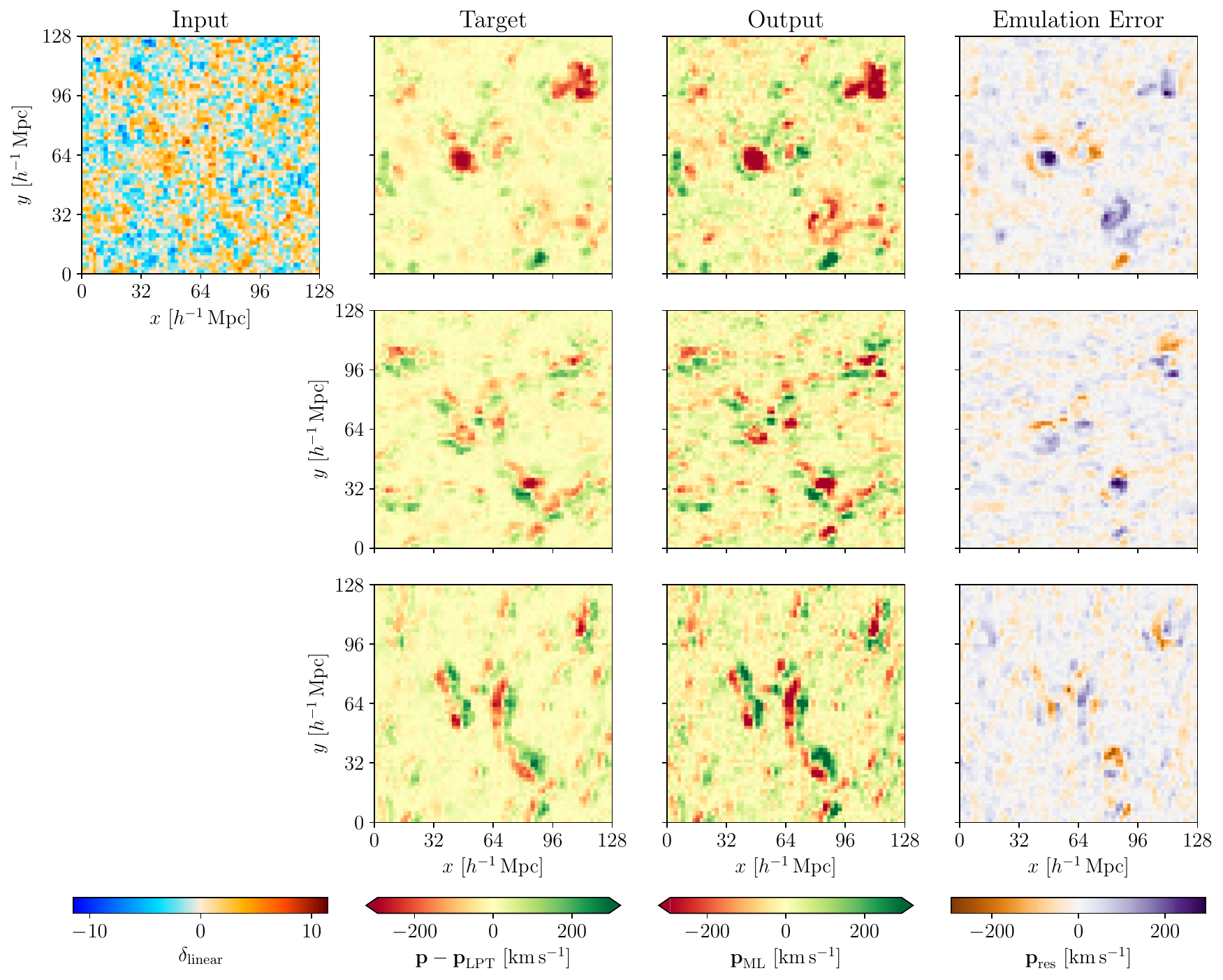}
    \caption{Slices of the input, target, output, and error of the frame of reference emulator at the final time step (i.e., with style parameter $a=1$). The input is the (scalar) linear density field (first column). The emulator aims to predict the three components (one per row) of its target $\textbf{p} - \textbf{p}_{\rm LPT}$ (second column). The emulator's predictions are shown in the third column, and the emulation error $\textbf{p}_\mathrm{res} = \textbf{p}-\textbf{p}_\mathrm{LPT}-\textbf{p}_\mathrm{ML}$ is shown in the final column. 
    }
    \label{fig:pres_slice}
\end{figure*}

In \cref{fig:pres_slice}, we plot slices of the input $\delta_\mathrm{linear}$, output $\textbf{p}_\mathrm{ML}$, target $\textbf{p}-\textbf{p}_\mathrm{LPT}$, and emulation error $\textbf{p}_\mathrm{res} \equiv \textbf{p}-\textbf{p}_\mathrm{LPT}-\textbf{p}_\mathrm{ML}$ for one of our test simulations, where all fields are evaluated at $a=1$. 
The target fields are obtained by running COLA simulations with 20 time steps, using initial conditions matching those of the test simulations, and saving the residuals between the calculated and LPT momenta.
As described in \cref{sec:architecture}, the input is the linear density field, comprising a single channel, whereas the output prediction is a three-component vector for each Lagrangian grid point. Since we are learning the residuals between the true momentum and the LPT prediction, correlations observed in $\textbf{p}-\textbf{p}_\mathrm{LPT}$  are highly localised, reflecting the accurate capture of large-scale modes by LPT.

Visually, there is a notable correlation between $\textbf{p} - \textbf{p}_\mathrm{LPT}$ (second column of \cref{fig:pres_slice}) and $\textbf{p}_{\rm ML}$ (third column of \cref{fig:pres_slice}). Leveraging the linear density field and scale factor information, the emulator accurately identifies the spatial structure of the $\textbf{p}_{\rm ML}$ field. The small emulation errors indicate its capability to predict magnitudes as well. We observe that the emulation error is signal-dependent, resulting in larger values of $\textbf{p}_{\rm res}$ in the regions where $\left|\textbf{p}_{\rm ML} \right|$ is large. These regions are highly nonlinear and appear as the simulation progresses. It is noteworthy that the emulation errors become particularly visible when visualising the final quantities in \cref{fig:pres_slice}, given their lesser prevalence at earlier times.

\begin{figure}
    \centering
    \includegraphics[width=\columnwidth]{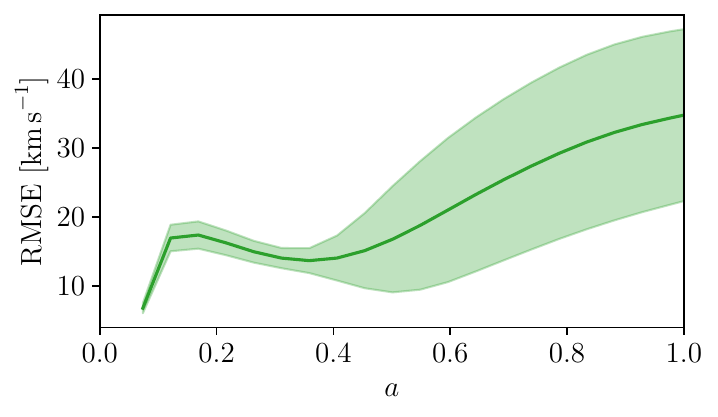}
    \caption{Root mean squared error (RMSE) of the emulated frame of reference, $\textbf{p}_{\rm ML}$, compared to $\textbf{p} - \textbf{p}_{\rm LPT}$, as a function of the scale factor $a$. The solid line indicates the mean across the test simulations, while the shaded band represents the standard deviation.
    }
    \label{fig:emulator_error}
\end{figure}

To quantify the magnitude and time-dependence of the emulation error, we plot the root mean squared error (RMSE) between the true $\textbf{p} - \textbf{p}_{\rm LPT}$ and $\textbf{p}_{\rm ML}$, as predicted by the emulator, as a function of the scale factor in \cref{fig:emulator_error}.
We present the mean and standard deviation of the RMSE across 50 test simulations.
At early times, the trajectories of the particles are well described by perturbation theory. Thus, even though LPT is not a perfect description of the dynamics, the emulator can easily correct for the error, maintaining a relatively constant RMSE of less than $20 {\rm \, km \, s^{-1}}$ for $a < 0.5$.
We observe a slight decrease in RMSE between $a=0.2$ and $a=0.4$, which is understandable given \cref{fig:time_dependent_pres}: initially, the field exhibits a high degree of small-scale structure, which becomes less significant and approximately constant over time, making $\textbf{p} - \textbf{p}_{\rm LPT}$ easier to predict during this period.
Beyond $a \approx 0.5$, the small-scale dynamics become highly non-linear, making it more challenging for the emulator to predict the correct frame of reference.
Consequently, we observe the behaviour schematically illustrated in \cref{fig:coca_sketch}: the emulation error grows at late times, approximately doubling between $a=0.5$ and $a=1$. 
\citet{Jamieson2023} and \citet{Jamieson2024} found similar issues with predicting virialised motions within collapsed regions due to their chaotic and random nature. It is precisely these emulation errors that we aim to correct using the COCA framework.

\subsection{COCA performance}
\label{sec:COCA_performance}

\begin{figure*}
    \centering
    \includegraphics[width=\textwidth]{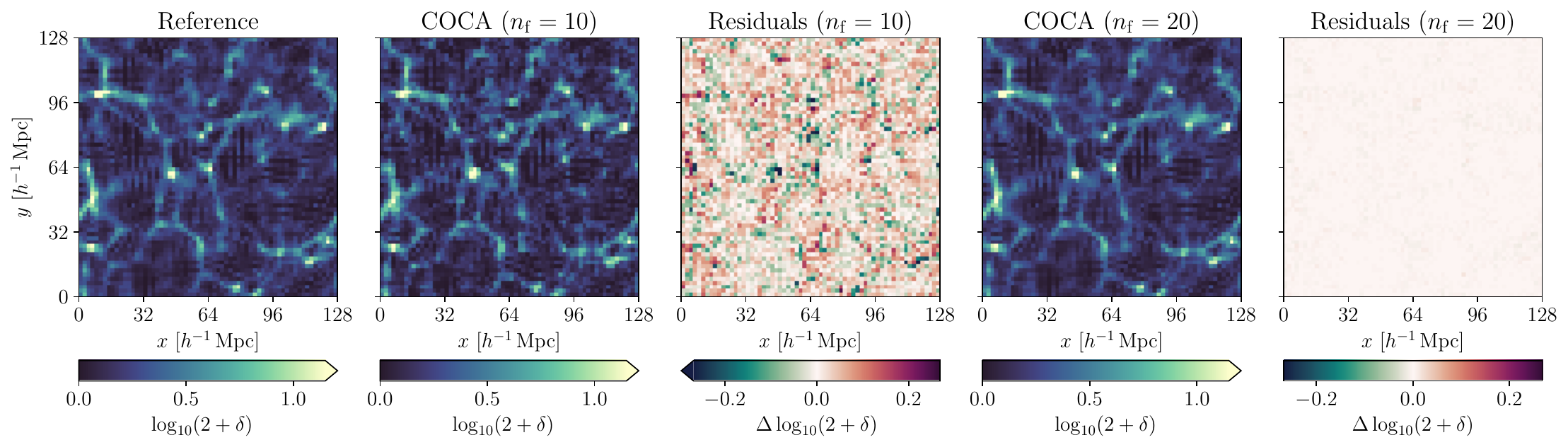}
    \caption{Slices of the final ($a=1$) matter density field for a reference simulation (first column) compared to the corresponding COCA simulations using either 10 (second column) or 20 (fourth column) force evaluations. The residuals relative to the reference simulation are shown in the third and fifth columns.}
    \label{fig:density_slice}
\end{figure*}

We now turn to testing the use of our frame of reference emulator within a cosmological simulation. To do this, for each realisation of initial conditions in our test set, we run a reference simulation (see \cref{sec:training_data}) as well as COCA and COLA simulations with varying specifications. For these runs, we use 20 time steps between $a=0.05$ and $a=1$, spaced linearly in scale factor, but we vary the number of force evaluations $n_{\rm f}$.
After some experimentation, we found that the best strategy to maximise the statistics described in \cref{sec:metrics} is to place all force evaluations at the end of the simulation.
This is expected, as the dynamics become more non-linear at later times, making it crucial to accurately resolve particle trajectories during these periods, especially since the emulation error is also largest at these times (see \cref{fig:emulator_error}).

In \cref{fig:density_slice}, we plot a slice of the final density field for one of the reference simulations in our test set, as well as the corresponding COCA simulations with $n_{\rm f} = 10$ and $n_{\rm f} = 20$, and their respective residuals relative to the reference. Both COCA simulations accurately recover the overall structure of the cosmic web, with correctly positioned filaments and nodes. With the smaller number of force evaluations, there is a small residual in the final density, but this has almost completely disappeared when $n_{\rm f} = 20$.

\begin{figure*}
    \centering
    \includegraphics[width=\textwidth]{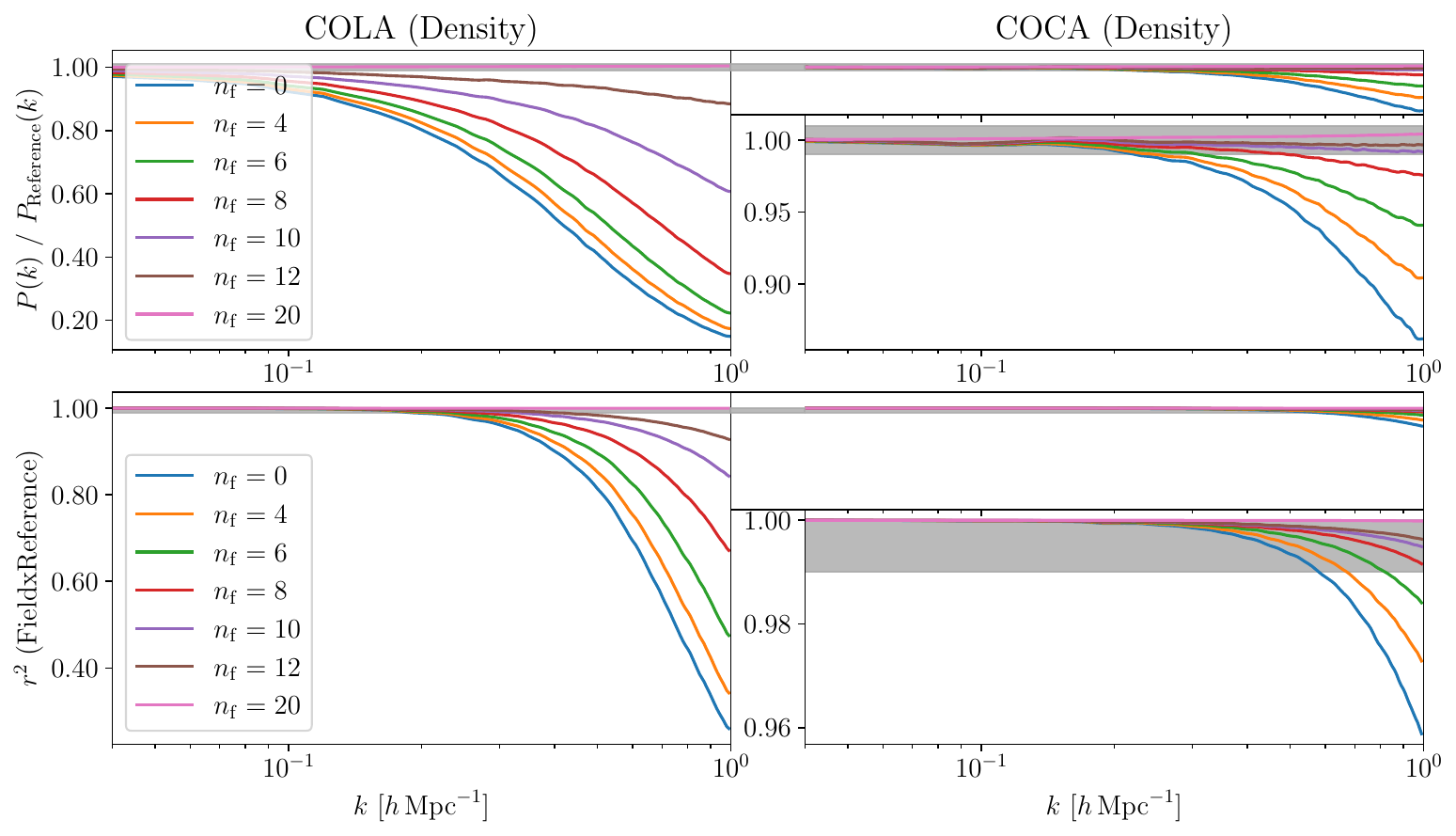}
    \caption{Ratio of the matter power spectrum (top row) and the cross-correlation (bottom row) with respect to the reference simulation, for COLA (left column) and COCA (right column) simulations with varying numbers of force evaluations, $n_{\rm f}$.
    The coloured lines represent the mean over the test set.
    In the COCA column, the top panel of each row is plotted on the same scale as COLA, while the lower panel provides a zoomed-in version. The grey band indicates 1\% agreement with the reference. COCA simulations are much closer to the reference even when using far fewer force evaluations, and the agreement improves as $n_{\rm f}$ increases.
    }
    \label{fig:timestepping_density}
\end{figure*}

To assess the relative performance of COCA and COLA and to determine the optimal number of force evaluations, in \cref{fig:timestepping_density} we plot the fractional error on the matter power spectra and the cross-correlation coefficient for the $a=1$ density field as a function of wavenumber for both simulation frameworks on the test set.
As a sanity check, we verify that both COLA and COCA achieve similar performance when performing a force evaluation at each of the 20 time steps.
Our first observation is that COCA performs dramatically better than COLA, even when using few force evaluations. It is unsurprising that with $n_{\rm f}=0$ COLA performs poorly, as this is merely the LPT prediction, which is known to be a poor description at this redshift and on these scales. In contrast, we find that COCA with $n_{\rm f}=0$ is already extremely accurate: purely following the trajectories of the emulated frame of reference ($n_{\rm f}=0$) produces a behaviour practically identical to running a COLA simulation with $n_{\rm f}=12$ force evaluations. The matter power spectrum of the emulated field is 99\% accurate up to $k \approx 0.3 \, h \, {\rm Mpc}^{-1}$, with $r^2(k) > 0.99$ up to $k \approx 0.6 \, h \, {\rm Mpc}^{-1}$. 
One would expect that, if the training simulations and evolution used a higher-accuracy gravity solver (e.g. P$^3$M or a tree-based approach), COCA would outperform COLA. However, it is not possible to check this conjecture in this example, since both the frame of reference emulator and COCA solver are based on PM forces.
Despite the good predictions of the emulator, we see that the relative error on the power spectrum increases to more than 10\% at $k=1 \, h \, {\rm Mpc}^{-1}$ when $n_\mathrm{f}=0$.
However, the error is reduced to less than 1\% up to $k = 0.5 \, h \, {\rm Mpc}^{-1}$ by adding just 8 force evaluations, and less than 1\% up to $k = 1 \, h \, {\rm Mpc}^{-1}$ with 10 force evaluations, both in terms of the power spectrum and phase accuracy.
This feature highlights the benefit of the COCA framework: we use machine learning to provide good approximations to the true solution and can run a physical simulation to correct for any errors made, using far fewer force evaluations than is ordinarily required.

\begin{figure*}
    \centering
    \includegraphics[width=\textwidth]{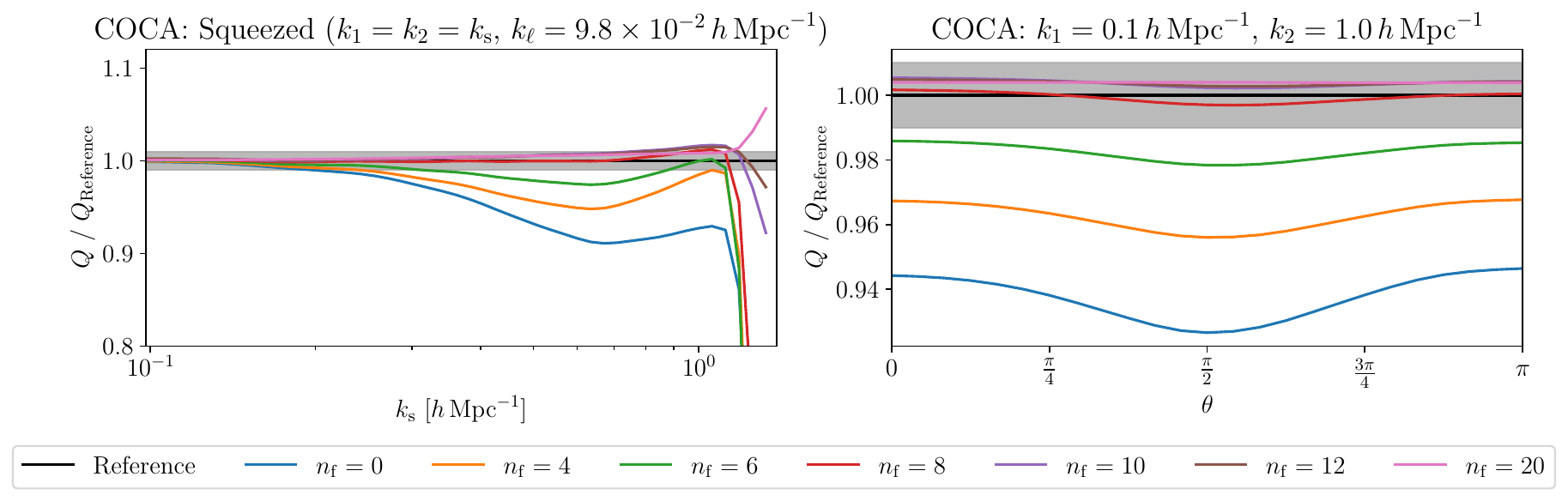}
    \caption{Ratio of the reduced bispectrum (\cref{eq:reduced_bispectrum}) between COCA simulations and the reference, as a function of the number of force evaluations, $n_{\rm f}$.
    In the left panel, we plot a squeezed configuration, where the third wavenumber is $k_3 = k_\ell = 9.8 \times 10^{-2} \, h \, {\rm Mpc}^{-1}$ and we vary $k_1 = k_2 = k_{\rm s}$.   
    In the right panel, we choose $k_1 = 0.1 \, h \, {\rm Mpc}^{-1}$ and $k_2 = 1.0 \, h \, {\rm Mpc}^{-1}$, and plot $Q$ as a function of the angle between these two vectors, $\theta$.
    The coloured lines represent the mean over the test set.
    The grey band indicates 1\% agreement.
    }
    \label{fig:bispectrum}
\end{figure*}

The same behaviour is observed when considering the three-point statistics. In \cref{fig:bispectrum} we plot the bispectrum for the COCA simulations in the configurations outlined in \cref{sec:metrics}. As with the power spectrum, reasonable agreement with the reference is achieved without any force corrections, with errors of the order of 5-10\%. However, with just 8 force evaluations, one achieves close to perfect agreement with the reference for almost all configurations considered, with the only discrepancy occurring for $k_{\rm s} > 1 \, h \, {\rm Mpc}^{-1}$.

\begin{figure*}
    \centering
    \includegraphics[width=\textwidth]{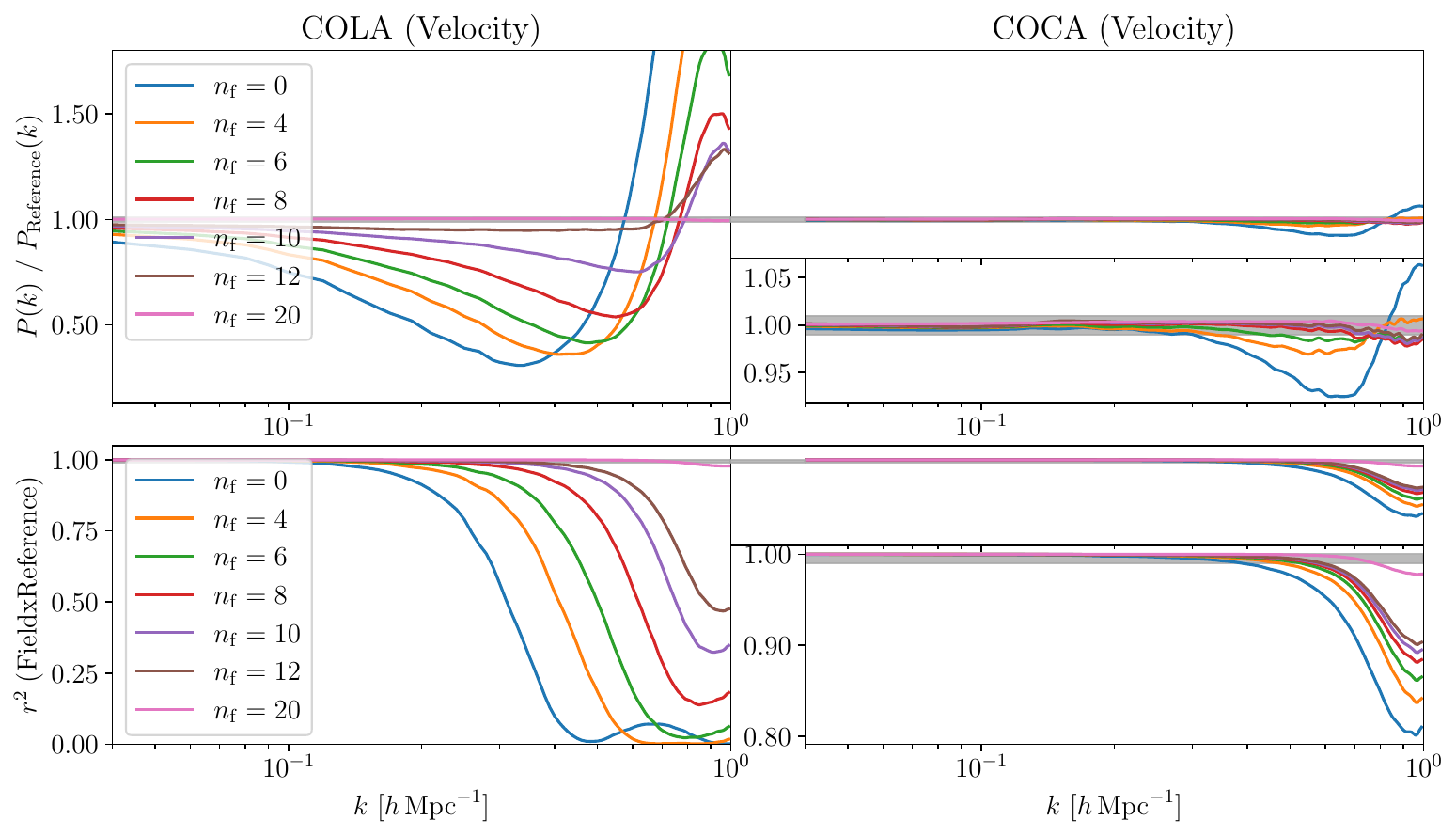}
    \caption{Same as \cref{fig:timestepping_density}, but for velocity field two-point statistics. Again, we see that COCA performs much better than COLA with fewer force evaluations, and that the result converges to the truth as $n_{\rm f}$ increases.
    }
    \label{fig:timestepping_velocity}
\end{figure*}

We now evaluate the accuracy of the simulated velocity fields by plotting the error on the power spectrum and cross-correlation coefficient for the final velocity potential in \cref{fig:timestepping_velocity}.
Velocity fields are very poorly predicted for all COLA simulations that skip force evaluations, with an under-prediction of power beyond $k \approx 0.1\, h \, {\rm Mpc}^{-1}$, and an over-prediction as one approaches the Nyquist frequency of our simulations.
The cross-correlation between the COLA velocities and the reference is also very low, with practically zero correlation at $k = 1 \, h \, {\rm Mpc}^{-1}$ when no force evaluations are used, and with $r^2(k) \approx 0.5$ at this scale for $n_{\rm f} = 12$.
This is unsurprising, since this latter case is equivalent to initialising a COLA simulation with an LPT prediction at a redshift of $z=1.5$ and using 12 time steps; one would not expect the initial conditions of such a simulation to be reasonable, as this is well beyond the validity of LPT.
However, this problem is alleviated if one uses an emulated frame of reference. Using only the emulator ($n_{\rm f} = 0$) reduces the error on the velocity field power spectrum to approximately 5\% at a $a=1$, which, although still reasonably large, is much smaller than what is found with COLA with up to $n_{\rm f}=12$.
The advantage of the COCA framework is particularly evident when varying $n_{\rm f}$, as the addition of just 6 force evaluations practically eliminates this error, reducing it to 1\%.
Similarly, we find that the COCA fields are much more correlated with the reference, even when using far fewer force evaluations, with $r^2(k) > 0.8$ for all $k \lesssim 1 \, h \, {\rm Mpc}^{-1}$ and for any number of force evaluations. The degree of correlation improves as one increases $n_{\rm f}$.

In summary, we find that our emulator can reasonably recover the density and velocity fields even without any correction. However, emulation errors of up to $\mathcal{O}(10\%)$ remain, but these can be reduced to the sub-percent level with just 8 force evaluations. Thus, COCA is able to correct for mistakes made in the emulation of particle trajectories by running a simulation in the corresponding frame of reference.

\subsection{COCA with misspecified cosmological parameters}
\label{sec:ml-safety}

\begin{figure*}
    \centering
    \includegraphics[width=0.49\textwidth]{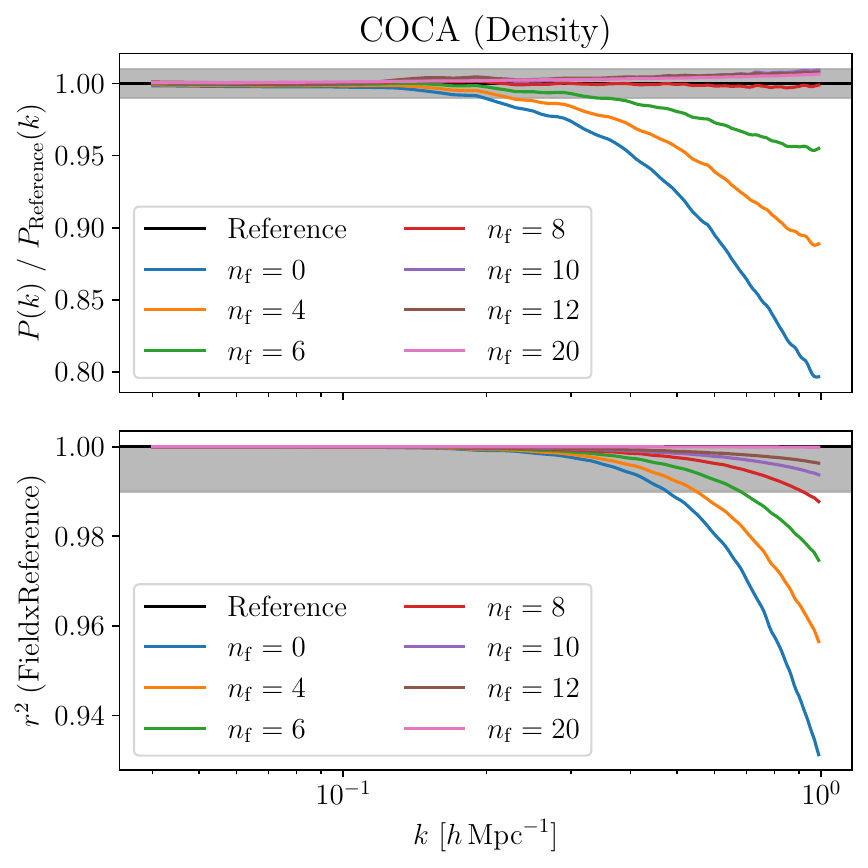}
    \includegraphics[width=0.49\textwidth]{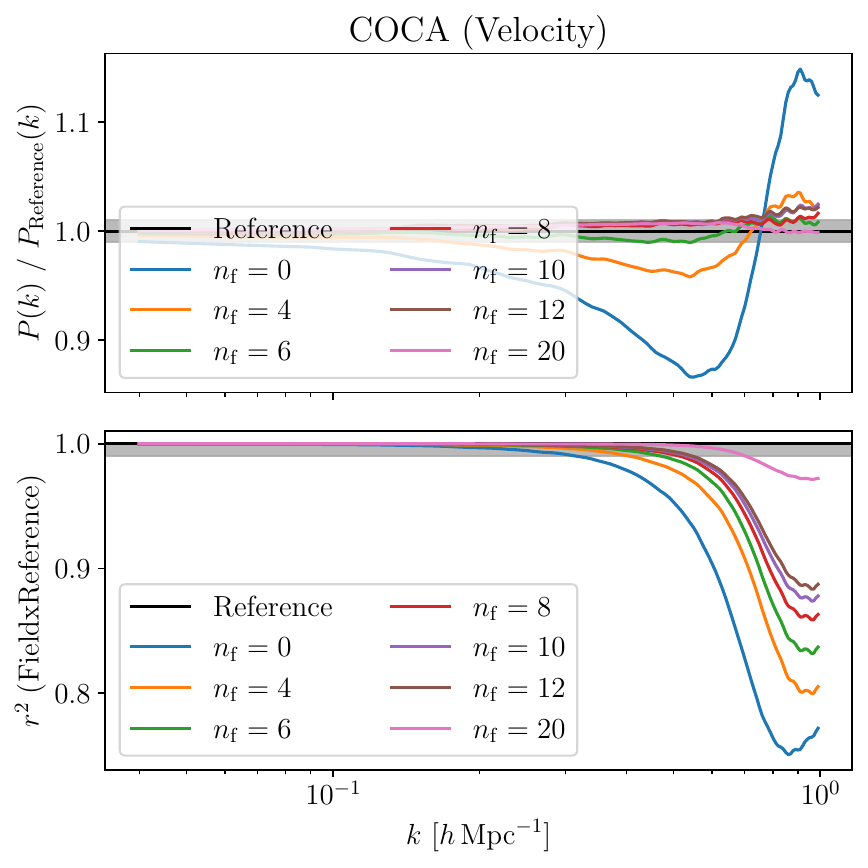}
    \caption{Performance of the COCA framework when applied outside the range of the training data. We compare the power spectrum (top row) and cross-correlation coefficient (bottom row) of the matter density field (left column) and velocity potential (right column) when using our emulator with a different set of cosmological parameters than it was trained on.
	The coloured lines show the mean across 50 test simulations as a function of the number of force evaluations $n_{\rm f}$. The grey band indicates 99\% agreement.
	In this test of robustness to cosmological parameter misspecification, only 8 force evaluations are required to correct the emulation error up to $k \gtrsim 0.6 \, h {\rm \, Mpc}^{-1}$.
    }
    \label{fig:cosmology_test}
\end{figure*}

One of the key motivations behind the COCA framework is the concept of ML-safety. Although emulation techniques have previously been applied to predict the results of dark matter simulations \citep{Perraudin2019,He2019,deOliveira2020,Jamieson2023,Jamieson2024,Giusarma2023,Conceicao2024,Saadeh2024}, there may be concerns that the emulated solutions might not match the truth if the initial conditions or cosmological parameters are ``unusual,'' i.e., unlike the training data.
The capacity of emulators to extrapolate was tested by \citet{Jamieson2022} in the context of well-understood simple matter distributions that had not been seen during training. Furthermore, \citet{Doeser2023} found that their emulator performed well with initial conditions containing significantly less power than their training examples.
However, with regular emulators, it is not possible to test all possible configurations, and thus, in general, one can only hope that the model extrapolates well to the application of interest.
In contrast, COCA uses a frame of reference emulator but solves the fundamental equations of motion. Therefore, any extrapolation mistake made in the emulation should be automatically corrected, unlike with the use of an emulator alone.

Our frame of reference emulator was trained using simulations run at a single cosmology. To test its out-of-distribution behaviour and its use in the COCA formalism, we ran 50 additional test simulations with a different set of cosmological parameters: $\Omega_{\rm b} = 0.03$, $\Omega_{\rm m} = 0.35$, $h = 0.7$, $n_{\rm s} = 0.99$, and $\sigma_8 = 0.9$.
These parameters are chosen to be relatively extreme, yet still within the support of a moderately wide prior that could be used for a cosmological analysis.
We note that we use the correct cosmological parameters for producing the initial density field, obtaining the LPT displacement fields, and solving the equations of motion; the only place where cosmological parameters are misspecified is in the prediction of $\textbf{p}_{\rm ML}$.

In \cref{fig:cosmology_test}, we plot the fractional error on the power spectra and the cross-correlation coefficients for the density and velocity fields in COCA simulations. The reference is the COLA simulations run with the same initial conditions, which are not subject to model misspecification in this scenario.
Despite the relatively extreme cosmological parameters, the uncorrected fields ($n_{\rm f}=0$) yield reasonable power spectra and cross-correlation. The mean error on the density power spectrum is approximately 20\% by $k=1 \, h \, {\rm Mpc}^{-1}$ with $r^2(k) > 0.93$ at these scales, while the velocity power spectrum has slightly smaller errors---around 10\%---and $r^2(k) \approx 0.85$ by $k=1 \, h \, {\rm Mpc}^{-1}$.
This moderate agreement with the truth is enabled by using the initial density field rather than the white noise field as the input to the emulator (see section \ref{sec:architecture}).
Indeed, even if the initial density appears different from that of the training simulations, the emulator does not have to predict the relevant initial matter power spectrum, which contains the entire dependence on all cosmological parameters except $\Omega_{\rm m}$.
Additionally, since one expects $\textbf{p}_{\rm ML}$ to be sourced only by local contributions in Lagrangian coordinates, the sensitivity to cosmological parameters should be relatively small.
Similar moderately accurate extrapolation behaviour has also been observed in other cosmological simulation emulators \citep{He2019,KodiRamanah2020,deOliveira2020,Lanzieri2022,Payot2023,Saadeh2024}.

Despite the moderate performance of this emulator in the presence of cosmological parameter misspecification, without any force evaluations (i.e., with $n_\mathrm{f} = 0$), the error on the matter power spectrum would be too large for current cosmological analyses \citep{Taylor2018}.
Therefore, relying solely on an emulator of particles' trajectories (i.e., a frame of reference emulator with $n_\mathrm{f} = 0$) as a forward model would produce inappropriate results and would not be a safe use of machine learning.
However, trajectories can be rectified in the COCA framework by evaluating gravitational forces and solving for the residual displacements with respect to the emulated frame of reference.
In our test, using just 8 force evaluations is sufficient to achieve percent-level agreement in both $P(k)$ and $r^2(k)$ for the density field, for all $k < 1 \, h \, {\rm Mpc}^{-1}$ (see \cref{fig:cosmology_test}). 
The same conclusion is true for the velocity field up to $k \approx 0.6 \, h \, {\rm Mpc}^{-1}$.

One might question whether an emulator trained on an incorrect cosmology could yield a worse frame of reference than 2LPT at the correct cosmology, thereby negating any advantage of COCA over COLA. We test this question in \cref{fig:x_emulator}, where we compare the performance of COCA using a reference frame derived from a misspecified cosmology (while all other computations, including gravitational evolution, employ the correct cosmology as noted above), against COLA, which uses the correct cosmological parameters.
For all summary statistics, we find that even with an incorrect cosmology, COCA produces substantially more accurate density fields for any given number of force evaluations. Thus, despite the presence of emulation inaccuracies in the misspecified scenario, our emulator still outperforms the 2LPT frame of reference and significantly reduces the number of force evaluations required to achieve an accurate density field. We have verified that this conclusion remains valid when varying $\Omega_{\rm m}$ within the range $[0.2, 0.4]$.

Thus, with only a small additional computational cost, we can convert an unsafe use of machine learning in cosmology into a well-behaved one, even when the emulator is applied outside the range of its training data. This is one of the main benefits of COCA compared to traditional emulators of $N$-body simulation results.

\subsection{COCA versus a Lagrangian displacement field emulator}

\begin{figure*}
    \centering
    \includegraphics[width=\textwidth]{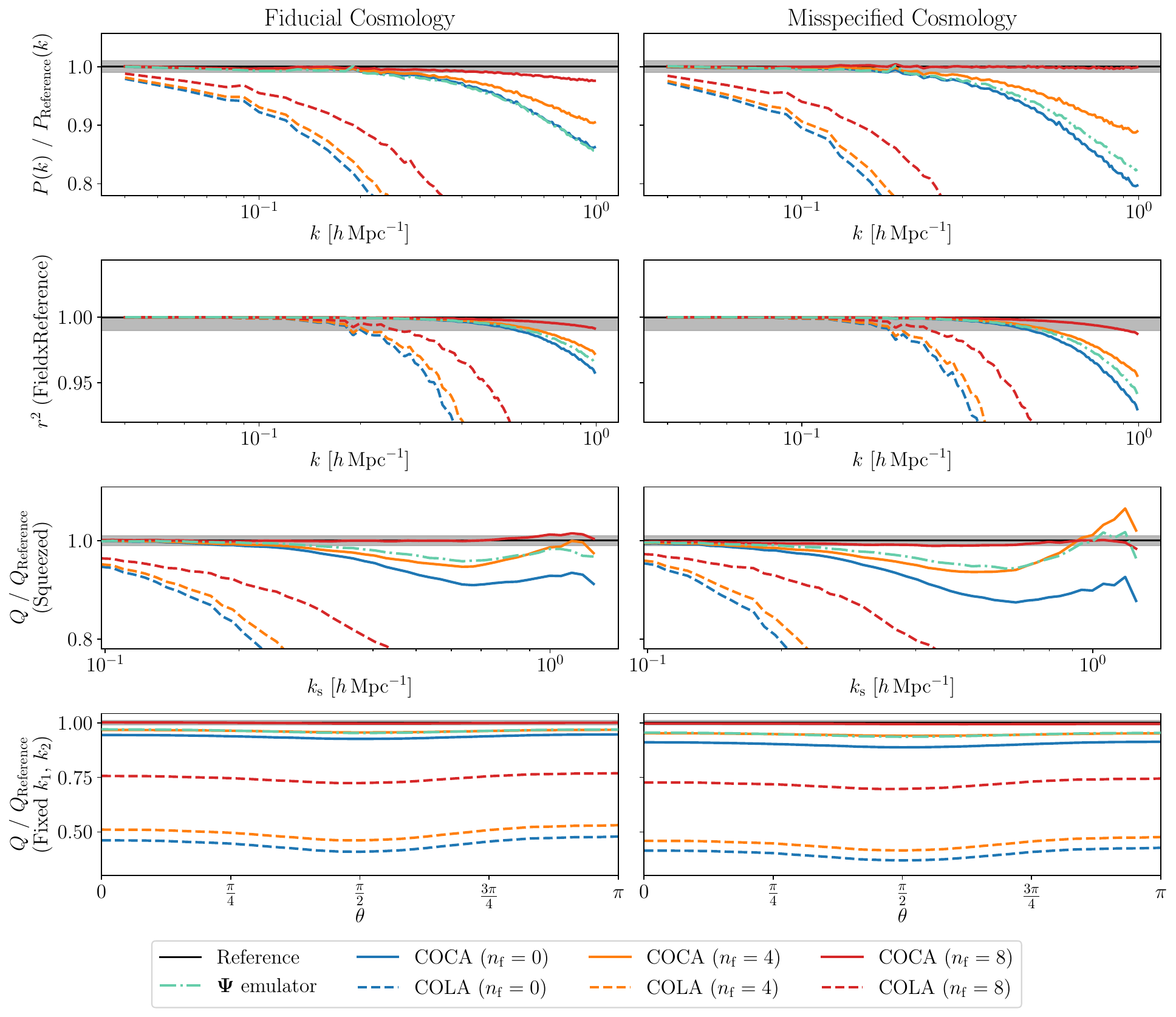}
    \caption{
    Relative performance of COCA and COLA versus an emulator of the displacement field $\boldsymbol{\Psi}$. We compute the summary statistics outlined in \cref{sec:metrics} for the final ($a=1$) matter density field and compare results at both the training cosmology and a misspecified one for COCA (COLA is evaluated at the correct cosmology in both cases). Although directly emulating $\boldsymbol{\Psi}$ produces a more accurate density field than simply emulating the momentum field $\textbf{p}$ (with $n_{\rm f} = 0$), using the COCA framework (emulating the frame of reference and employing additional force evaluations) yields the best performance.
    Moreover, even when using a misspecified cosmology to predict the frame of reference, COCA significantly outperforms COLA for any given number of time steps.
    }
    \label{fig:x_emulator}
\end{figure*}

In this work, we advocate for using an emulator for the frame of reference in $N$-body simulations, as it allows for correcting emulation errors by introducing force evaluations. This approach contrasts with previous emulators \citep[e.g.][]{He2019,deOliveira2020,Jamieson2023,Jamieson2024}, which directly predict the Lagrangian displacement field, i.e., the simulation output. This section compares the relative accuracy of these two approaches as a function of the number of force evaluations in COCA.

To investigate this question, we train a time-dependent emulator for the residual displacement field $\boldsymbol{\Psi}_{\rm ML} (\textbf{q}, a)$, defined as the difference between the true Lagrangian displacement field $\boldsymbol{\Psi} (\textbf{q}, a)$ and that predicted by LPT, $\boldsymbol{\Psi}_{\rm LPT} (\textbf{q}, a)$. We opted to train a new $\boldsymbol{\Psi}$-emulator rather than directly compare COCA to existing literature results to minimise the impact of differences in gravity solvers, training set sizes, architecture choices, and training procedures. For a fair comparison, we trained our $\boldsymbol{\Psi}$-emulator using the same simulations as for the frame of reference emulator, employing the same architecture and training procedure outlined in \cref{sec:training} (with $\textbf{p}$ replaced by $\boldsymbol{\Psi}$ in the loss function).

In a similar manner as before, we begin by normalising the target variable by defining the function $\psi(a)$ such that
\begin{equation}
    \boldsymbol{\Psi}_{\rm ML} (\textbf{q}, a) \equiv \psi(a) \tilde{\boldsymbol{\Psi}}_{\rm ML} (\textbf{q}, a),
\end{equation}
where $\tilde{\boldsymbol{\Psi}}_{\rm ML} (\textbf{q}, a)$ has unit standard deviation. Applying symbolic regression to the function $\psi(a)$, we find that it is well approximated by 
\begin{equation}
     \psi(a) \approx a^{\phi_0} + \phi_1,
\end{equation}
with $\phi_0 = 1.2412539$ and $\phi_1 = -0.05402543$, yielding a root mean squared error of $5 \times 10^{-4}$.
We take the linear density field as input, but this time output $\tilde{\boldsymbol{\Psi}}_{\rm ML} (\textbf{q}, a)$.
We evaluate this emulator on the same test simulations as for the frame of reference emulator, and convert the returned Lagrangian displacements into an Eulerian density field using a cloud-in-cell estimator.
We compute the power spectrum, cross-correlation coefficient, and bispectra (see \cref{sec:metrics}) and plot these in \cref{fig:x_emulator}, where we compare against COCA without force evaluations ($n_{\rm f}=0$; using solely the frame of reference emulator) and both $n_{\rm f}=4$ and $n_{\rm f}=8$.
We perform this analysis for both the fiducial and misspecified cosmology (see \cref{sec:ml-safety}).

When compared to COCA with $n_{\rm f} = 0$, the Lagrangian displacement field emulator more accurately recovers the reference density field. The power spectra of the two methods are relatively similar with fiducial cosmological parameters, but the difference becomes more pronounced when the cosmology is misspecified. For all other metrics, the $\boldsymbol{\Psi}$-emulator produces summary statistics that are closer to the reference.
This behaviour is expected: the $\boldsymbol{\Psi}$-emulator is designed to optimise the prediction of dark matter particle positions through its loss function, naturally resulting in an accurate density field. In contrast, the frame of reference emulator in COCA aims to match particle momenta $\textbf{p}$. Consequently, without force evaluations, emulation errors in $\textbf{p}$ accumulate over time, reducing the quality of the final density field.

Although the $\boldsymbol{\Psi}$-emulator performs better than COCA with no force evaluations, there is no way to correct its errors, meaning its performance cannot be improved. Conversely, in COCA, force evaluations can be added to correct the errors made by the frame of reference emulator.
\cref{fig:x_emulator} shows that the addition of only four force evaluations results in performance nearly identical to that of the $\boldsymbol{\Psi}$-emulator for the bispectra, and better results for two-point statistics with residual errors reduced by a factor of 1.4.
Residual errors almost entirely disappear when eight force evaluations are used in COCA: the final power spectrum $P(k)$ has approximately four to five times smaller errors than the one derived from the $\boldsymbol{\Psi}$-emulator at all scales for both cosmologies.
Thus, even with very limited additional computations beyond the emulation, the COCA framework outperforms a Lagrangian displacement field emulator.

We note that our displacement emulator is slightly less accurate than that of \citet{He2019}, who also emulated a PM-like output. They achieved errors on $P(k)$ of 0.8\% and 4\% at $k=0.4 \, h {\rm \, Mpc}^{-1}$ and $k=0.7 \, h {\rm \, Mpc}^{-1}$, respectively, whereas our emulator is accurate to 3\% and 9\% at these scales.
We attribute this discrepancy to our use of fewer training simulations (2,000 fields compared to 10,000 in \citealp{He2019}), the need for our emulator to learn time-dependence (only 100 of the 2,000 training fields are at $a=1$), and \citet{He2019} employing a more optimised architecture and training schedule.
As mentioned in \cref{sec:training_data}, since the aim of this paper is to demonstrate how to correct for emulation errors rather than produce the optimal emulator, we chose not to increase the number of training simulations or fine-tune the architecture, as our emulator is already of similar quality to those in the literature.
If a frame of reference emulator with performance similar to that of \citet{He2019} were used in COCA, fewer time steps would be needed to correct for emulation errors, thereby achieving the same theoretical guarantees with reduced computational expense.

\subsection{Timing tests}

\begin{figure}
    \centering
    \includegraphics[width=\columnwidth]{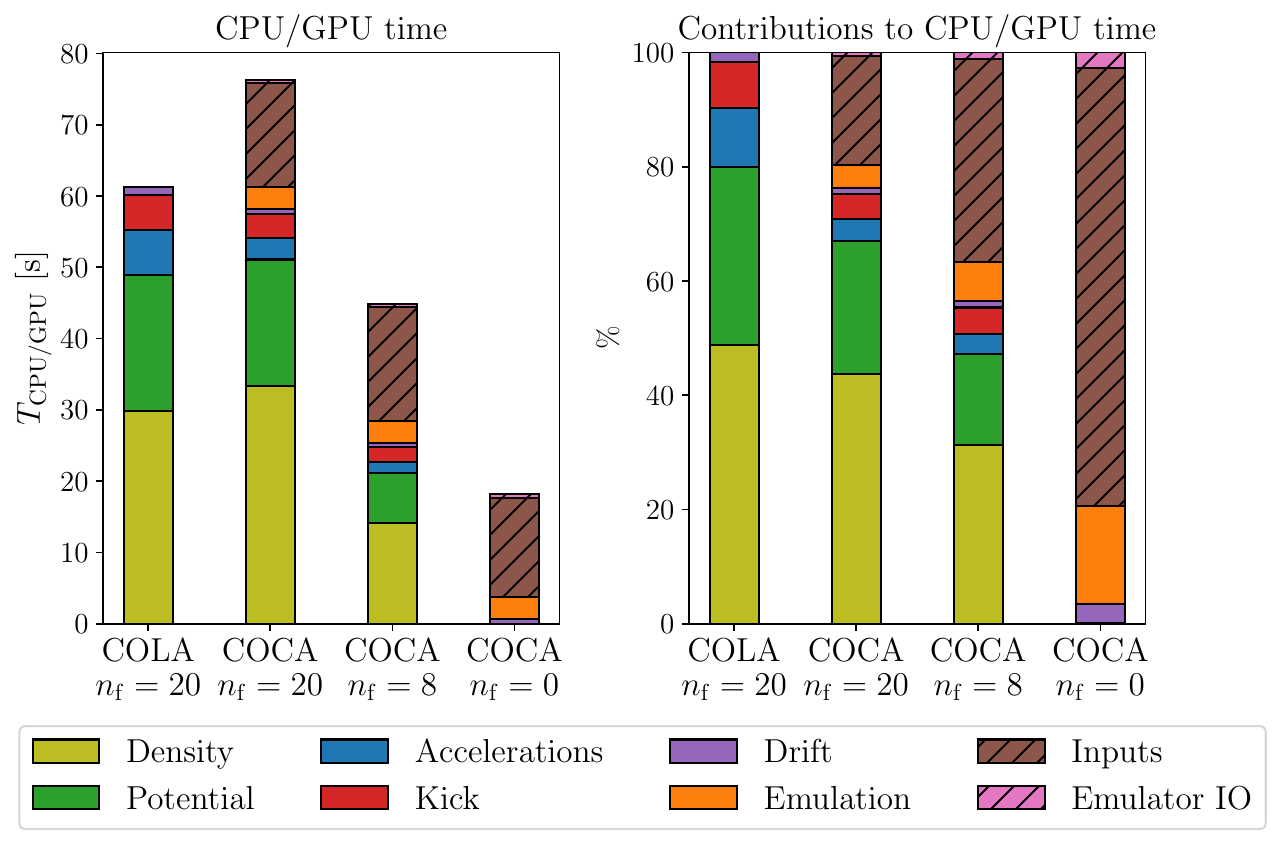}
    \caption{Total CPU/GPU time for COCA with varying numbers of force evaluations, $n_{\rm f}$, compared to COLA. Both COCA and COLA use 20 time steps. In the left column, we report the total time, and in the right column, we report the relative contributions of the various operations. Solid bars correspond to the main computations, and hashed bars indicate inputs and outputs. 
We note that the current implementation of COCA in \sbmy{} decouples emulation and other computations, which is not a requirement---emulation could occur on the fly during $N$-body evolution. In this way, the cost of input/output operations would be effectively reduced to zero.
    }
    \label{fig:speed}
\end{figure}

To assess the computational performance of COCA, in \cref{fig:speed} we show the required amount of CPU/GPU time for each of the stages of the framework. To perform the timing tests, we use an Intel Xeon Gold 6230 processor with 40 CPU cores and an Nvidia V100 GPU. We compare the results of running COLA with 20 time steps (and 20 force evaluations) and COCA with the same number of time steps, but with varying numbers of force evaluations (0, 8, and 20). All other settings are identical to those in section \ref{sec:COCA_performance}.

The COLA simulation takes approximately 61 CPU-seconds to run, with almost half of this time spent on cloud-in-cell binning (converting particle positions to the density field).
In our test, running COCA with $n_{\rm f}=0$ is approximately four times faster. 
In the current implementation, the emulation and the simulation codes are disjoint, with the frame of reference being written to and then read from disk at each kick time step (a process responsible for 77\% of the CPU/GPU time in this case).
Separating emulation and $N$-body evolution is not a fundamental requirement of the COCA framework: one could emulate on the fly, which would effectively reduce the input/output time to zero for a slightly higher memory cost (two $\textbf{p}_\mathrm{ML}$ fields need to be kept in memory for each kick operation, see \cref{eq:ML_acceleration}.
Such an approach would make the $n_{\rm f}=0$ case 18 times faster than the COLA simulation, with this emulator.

To enable the safe use of an ML emulator, COCA relies on including a finite number of force evaluations.
Naturally, if one uses the same number of force evaluations, then COCA is more computationally expensive than COLA, since it must perform the same steps as COLA but with an additional emulation stage.
However, because the ML correction makes the frame of reference more accurate than LPT alone, the number of force evaluations can be reduced to approximately 8 (see \cref{sec:COCA_performance}).
With the current implementation of separate emulation and $N$-body evolution codes, the cost of COCA with $n_\mathrm{f}=8$ is approximately two-thirds of the cost of the COLA simulation. If emulation were done on the fly, the time required for inputs (36\% of the total time) would be eliminated, making COCA 2.3 times faster than COLA.

These timing improvements are expected to become more dramatic if COCA were extended to include a more accurate gravity solver, such as a P$^3$M or tree-based code.
The computational expense for computing the forces in these codes is significantly higher than in the PM-based model used in this work. Therefore, reducing the number of force evaluations would dramatically improve run time. We leave such an investigation to future work.

\section{Discussion and Conclusion}
\label{sec:conclusions}

In this paper, we have introduced COmoving Computer Acceleration (COCA), a hybrid formalism involving ML and $N$-body simulations. Unlike previous works that directly emulate the simulation output, COCA solves the dynamics of an $N$-body simulation using a machine-learnt frame of reference.
COCA can be seen as an improvement of COLA, which solves the dynamics of an $N$-body simulation in the LPT frame of reference.
By virtue of the principle of Galilean invariance, equations of motion can be solved in any frame of reference, making COCA a ML-safe framework.
COCA is the first framework to use physics to determine the (otherwise uncorrected) emulation error in $N$-body simulations using ML and correct for it.

The concept behind COCA is entirely independent of the $N$-body solver and of the ML emulation algorithm used. For this proof-of-concept, we employed a PM approach to solving the equations of motion and a V-net architecture for the frame of reference, with fixed cosmological parameters.
We have demonstrated that after the ML-prediction of the optimal frame of reference (the one in which all particles are at rest), running the $N$-body simulation corrects for potential emulation errors in the particle trajectories.
We have quantitatively shown that the number of force evaluations required to achieve a given accuracy is reduced compared to COLA. The frame of reference emulator achieves between 1\% and 10\% accuracy when used in isolation, but only eight force evaluations are needed to reduce emulation errors to the percent level, compared to a 100-time step COLA simulation. Therefore, COCA can be utilised as a cheap $N$-body simulator.
Furthermore, with eight force evaluations, COCA is four to five times more accurate than a Lagrangian displacement field emulator, when the frame of reference emulator and the Lagrangian displacement field emulator are trained using the same computational resources. This increased accuracy is due to COCA's ability to correct emulation errors and represents one of the main advantages of this framework compared to the direct emulation approach explored in earlier literature.

In \cref{sec:ml-safety}, we demonstrated that our frame of reference emulator is moderately robust to changes in cosmological parameters (despite training at a fixed cosmology). However, the COCA framework can correct for extrapolation errors arising from applying the emulator outside the range of validity of the training simulations.
Even when the frame of reference is inaccurate (because the ML training/prediction and the $N$-body evolution use different cosmological parameters), we found that percent-level accuracy can be reached on final density and velocity fields up to $k \approx 0.6 \, h \, {\rm Mpc}^{-1}$.
Thus, the COCA framework provides ML-safety even when models are required to extrapolate.
There is no fundamental reason why the emulator cannot depend on cosmological parameters, and future implementations of COCA can include these as additional style parameters of the neural network.

Our example focused on relatively small simulation volumes (with a side length of $128~h^{-1} {\rm \, Mpc}$) compared to those required for modern-day surveys (typically several gigaparsecs in length). With the current memory limitations of GPUs, it is not possible to emulate the entire volume with a single emulator at the desired resolution.
As a workaround, \citet{Jamieson2023} splits the volume into several padded sub-boxes and treats each one separately, relying on sequential predictions for particle displacements in each sub-box to cover the full volume.
Similarly, in COCA, one could predict the frame of reference for particles in each sub-box, and then solve the equations of motion in each sub-box independently.
This idea relates to the algorithm introduced by \citet{Leclercq2020sCOLA} for perfectly parallel $N$-body simulations using spatial comoving Lagrangian Acceleration (sCOLA).
There, a tiling of the simulation volume is used, and the evolution of tiles is spatially decoupled by splitting the Lagrangian displacement field into large and small-scale contributions.
In sCOLA, the frame of reference used in the evolution of tiles is given by LPT, but it could be easily replaced by a frame of reference including both LPT and an ML contribution, as introduced in this paper.
Such an approach would overcome the memory limitations of GPUs, which currently limit COCA to small simulation volumes.
An additional benefit of this approach would be the inexpensive generation of light-cones. Indeed, when using a tiling approach as with sCOLA, only one tile needs to be evolved to a redshift of zero; the tiles farthest from the observer only need to be evolved until they intersect the light-cone at higher redshifts.

It is important to emphasise that the specific implementation details used in this work are not requirements but just an example.
For instance, one could use a perturbation theory-informed integrator for the equations of motion \citep{Feng2016,ListHahn2023}, an approach complementary to COLA for fast generation of approximate cosmological simulations.
Furthermore, instead of using training simulations run with a PM gravity solver, one could learn the frame of reference given simulations with higher force accuracy, for example using a P$^3$M or tree-based gravity solver.
Subsequently, solving the equations of motion in the emulated frame of reference with the same solver would result in simulations with similar accuracy to those of the training set, but with significantly reduced computational cost.
The guarantee that any emulation mistakes are removed asymptotically as the number of force evaluations increases---a central feature of COCA---will remain.
This ML-safety cannot be guaranteed through direct emulation of P$^3$M or tree-based simulations.
As with COLA, the COCA framework could be adapted to include more extended physical models, such as neutrinos, which induce a scale-dependent growth factor \citep{Wright2017}.
Finally, although in this work we have focused on gravitational $N$-body simulations in a cosmological context, the approach of solving equations of motion in an emulated frame of reference could be applied to any kind of simulation involving interacting particles (e.g., electrodynamics, hydrodynamics, radiative transfer, magnetohydrodynamics).
We generally expect a reduction in computational demands while retaining physical guarantees of convergence to the truth.

Benefiting from its modest computational cost, COCA could be used in analyses of cosmological data using fully non-linear models. It could straightforwardly be used as a forward model in implicit likelihood inference algorithms such as \textsc{delfi} \citep{Alsing2019,Makinen2021}, \textsc{bolfi} \citep{Leclercq2018}, \textsc{selfi} \citep{Leclercq2019,Leclercq2022}, or the \textsc{LtU-ili} pipeline \citep{Ho2024}.
As COCA is an ML-safe framework, its use as a forward model cannot bias the inference result.
We also note that using a V-net emulator and a PM force solver, the entire COCA framework is differentiable. For the emulation of the frame of reference, differentiability is achieved via automatic differentiation. For the $N$-body evolution, differentiable PM simulators already exist \citep{Wang2014,Jasche2019,Modi2021,Li2022}. Building upon these, future work could be dedicated to writing a differentiable COCA solver, which could be used in Bayesian large-scale structure inference using an explicit field-level likelihood \citep[see][]{Jasche2019,Doeser2023,Wempe2024}.

Machine learning offers great promise in the acceleration of forward modelling in the physical sciences.
The output of any ML model is usually an approximation with inevitable emulation errors. In this paper, we have shown that emulation errors are correctable in gravitational $N$-body simulations.
By solving the correct physical equations while using the ML solution as an approximation, one can exploit the speed of ML while retaining the safety of more traditional methods.  

\onecolumngrid

\appendix

\section{The actual equations}
\label{apx:actual equations}

\subsection{Model equations with COCA}

Using the notations of \citet[][appendix B]{LeclercqThesis} and \citet{Leclercq2020sCOLA}, we consider dark matter particles with positions $\textbf{x}$ and momenta $\textbf{p}$ in comoving coordinates. Denoting the scale factor as $a$ and the over-density field as $\delta$, the equations to be solved are:
\begin{eqnarray}
\deriv{\textbf{x}}{a} & = & \mathpzc{D}(a) \textbf{p} \quad \mathrm{with} \quad \mathpzc{D}(a) \equiv \frac{1}{a^2 \mathcal{H}(a)}, \label{eq:drift}\\
\deriv{\textbf{p}}{a} & = & \mathpzc{K}(a) \boldsymbol{\nabla} \left( \Delta^{-1} \delta \right) \quad \mathrm{with} \quad \mathpzc{K}(a) \equiv -\frac{3}{2} \frac{\Omega_\mathrm{m}^{(0)} \mathcal{H}^{(0)2}}{a \mathcal{H}(a)},\label{eq:kick}
\end{eqnarray}
where $\mathcal{H}(a) \equiv a^\prime / a$ is the conformal Hubble factor (where a prime denotes a derivative with respect to conformal time) and $\Omega_\mathrm{m}^{(0)}$ is the matter density parameter at the present time ($a=1$). For simplicity, we note $\boldsymbol{\nabla}_\textbf{x} = \boldsymbol{\nabla}$, $\Delta_\textbf{x} = \Delta$ and $\delta(\textbf{x},a) = \delta$.

With $\textbf{x}(a) = \textbf{x}_\mathrm{LPT}(a) + \textbf{x}_\mathrm{ML}(a) + \textbf{x}_\mathrm{res}(a)$ (denoting the Lagrangian perturbation theory, machine-learnt, and residual contributions to the position, respectively), 
we note for each contribution $\mathrm{y} \in \left\lbrace \mathrm{LPT}, \mathrm{ML}, \mathrm{res} \right\rbrace$:
\begin{equation}
\deriv{\textbf{x}_\mathrm{y}}{a} \equiv \mathpzc{D}(a) \textbf{p}_\mathrm{y} \quad \mathrm{and} \quad \deriv{\textbf{p}_\mathrm{y}}{a} = \frac{\drm}{\drm a} \left( \frac{1}{\mathpzc{D}(a)} \deriv{\textbf{x}_\mathrm{y}}{a} \right) \equiv - \mathpzc{K}(a) \mathpzc{V}[\textbf{x}_\mathrm{y}](a),
\end{equation}
where the differential operator $\mathpzc{V}[\cdot](a)$ is defined by
\begin{equation}
\mathpzc{V}[\cdot](a) \equiv - \frac{1}{\mathpzc{K}(a)} \frac{\drm}{\drm a} \left( \frac{1}{\mathpzc{D}(a)} \deriv{\,\cdot}{a} \right) .
\end{equation}
Analogously, one writes the momenta as $\textbf{p}(a) = \textbf{p}_\mathrm{LPT}(a) + \textbf{p}_\mathrm{ML}(a) + \textbf{p}_\mathrm{res}(a)$, and thus Eqs. \eqref{eq:drift} and \eqref{eq:kick} take the form
\begin{eqnarray}
\deriv{\textbf{x}}{a} & = & \mathpzc{D}(a) \left\lbrace \textbf{p}_\mathrm{res}(a) + \textbf{p}_\mathrm{LPT}(a) + \textbf{p}_\mathrm{ML}(a) \right\rbrace, \\
\deriv{\textbf{p}_\mathrm{res}}{a} & = & \mathpzc{K}(a) \left\lbrace  \left[ \boldsymbol{\nabla} \left( \Delta^{-1} \delta \right) \right]\!\!(a) + \mathpzc{V}[\textbf{x}_\mathrm{LPT}](a) +  \mathpzc{V}[\textbf{x}_\mathrm{ML}](a) \right\rbrace. 
\end{eqnarray}

The analytical properties of LPT are \citep[see e.g.][equations (1.7), (1.96), (1.118) and appendix B]{LeclercqThesis}:
\begin{eqnarray}
\textbf{x}_\mathrm{LPT}(a) & = & \textbf{q} - D_1(a) \boldsymbol{\Psi}_1 + D_2(a) \boldsymbol{\Psi}_2 , \label{eq:LPT-mapping} \\
\mathpzc{D}(a) \textbf{p}_\mathrm{LPT} & = & - \deriv{D_1}{a} \boldsymbol{\Psi}_1 + \deriv{D_2}{a} \boldsymbol{\Psi}_2 , \\
\mathpzc{V}[\textbf{x}_\mathrm{LPT}](a) & = & - D_1(a) \boldsymbol{\Psi}_1 + \left[D_2(a) - D_1^2(a)\right] \boldsymbol{\Psi}_2 ,
\end{eqnarray}
where $\boldsymbol{\Psi}_1$ and $\boldsymbol{\Psi}_2$ are the time-independent first and second order displacements, with corresponding growth factors $D_1$ and $D_2$.
This gives
\begin{eqnarray}
\deriv{\textbf{x}}{a} & = & \mathpzc{D}(a) \left\lbrace \textbf{p}_\mathrm{res}(a) + \textbf{p}_\mathrm{ML}(a) \right\rbrace - \deriv{D_1}{a} \boldsymbol{\Psi}_1 + \deriv{D_2}{a} \boldsymbol{\Psi}_2, \label{eq:drift-COCA-standard}\\
\deriv{\textbf{p}_\mathrm{res}}{a} & = & \mathpzc{K}(a) \left\lbrace  \left[ \boldsymbol{\nabla} \left( \Delta^{-1} \delta \right) \right]\!\!(a) - D_1(a) \boldsymbol{\Psi}_1 + \left[ D_2(a) - D_1^2(a)\right]  \boldsymbol{\Psi}_2 + \mathpzc{V}[\textbf{x}_\mathrm{ML}](a) \right\rbrace. \label{eq:kick-COCA-standard}
\end{eqnarray}

Furthermore, for any arbitrary positive function $u$ of $a$, we can rewrite
\begin{eqnarray}
\deriv{\textbf{x}}{a} & = & \mathpzc{D}(a) u(a) \left\lbrace \frac{1}{u(a)} \times \textbf{p}_\mathrm{res}(a) + \frac{1}{u(a)} \times \textbf{p}_\mathrm{ML}(a) \right\rbrace - \deriv{D_1}{a} \boldsymbol{\Psi}_1 + \deriv{D_2}{a} \boldsymbol{\Psi}_2, \label{eq:drift-COCA-modified}\\
\deriv{\textbf{p}_\mathrm{res}}{a} & = & \deriv{u(a)}{a} \left\lbrace \frac{\mathpzc{K}(a)}{\drm u(a)/\drm a} \times \left[ \left[ \boldsymbol{\nabla} \left( \Delta^{-1} \delta \right) \right]\!\!(a) - D_1(a) \boldsymbol{\Psi}_1 + \left[ D_2(a) - D_1^2(a) \right] \boldsymbol{\Psi}_2 + \mathpzc{V}[\textbf{x}_\mathrm{ML}](a) \right] \right\rbrace. \label{eq:kick-COCA-modified}
\end{eqnarray}

\subsection{Time stepping with COCA}

In this paper, we adopt the second order symplectic ``kick-drift-kick'' algorithm, also known as the leapfrog scheme \citep[e.g.][]{Birdsall1985}, to integrate the equations of motion, for a series of $n+1$ time steps $t(a)$ between $t_0=t(a_\mathrm{i})$ and $t_{n+1}=t(a_\mathrm{f})$. This algorithm relies on integrating the model equations on small time steps and approximating the momenta and accelerations that appear in the integrands (the part between curly brackets in the model equations) by their value at some time within the interval.

The discrete versions of the COCA model equations (equations \eqref{eq:drift-COCA-standard}--\eqref{eq:kick-COCA-standard} or \eqref{eq:drift-COCA-modified}--\eqref{eq:kick-COCA-modified}) give the Drift and Kick operators for COCA:
\begin{eqnarray}
\label{eq:D_COCA}
\mathrm{D}(t_\mathrm{i}^\mathrm{D},t_\mathrm{f}^\mathrm{D},t^\mathrm{K}) : & \quad & \textbf{x}(t_\mathrm{i}^\mathrm{D}) \mapsto \textbf{x}(t_\mathrm{f}^\mathrm{D}) = \textbf{x}(t_\mathrm{i}^\mathrm{D}) + \alpha_\textbf{p}(t_\mathrm{i}^\mathrm{D}, t_\mathrm{f}^\mathrm{D}, t^\mathrm{K}) \textbf{p}_\mathrm{res}\!\left(t^\mathrm{K}\right) - \left[ D_1 \right]_{t_\mathrm{i}^\mathrm{D}}^{t_\mathrm{f}^\mathrm{D}} \boldsymbol{\Psi}_1 + \left[ D_2 \right]_{t_\mathrm{i}^\mathrm{D}}^{t_\mathrm{f}^\mathrm{D}} \boldsymbol{\Psi}_2 \\
& & \hfill \quad\quad\quad\quad\quad\quad\quad + \,\, \alpha_\textbf{p}(t_\mathrm{i}^\mathrm{D}, t_\mathrm{f}^\mathrm{D}, t^\mathrm{K}) \textbf{p}_\mathrm{ML}\!\left(t^\mathrm{K}\right), \nonumber \\
\mathrm{K}(t_\mathrm{i}^\mathrm{K},t_\mathrm{f}^\mathrm{K},t^\mathrm{D}) : & \quad & \textbf{p}_\mathrm{res}(t_\mathrm{i}^\mathrm{K}) \mapsto \textbf{p}_\mathrm{res}(t_\mathrm{f}^\mathrm{K}) = \textbf{p}_\mathrm{res}(t_\mathrm{i}^\mathrm{K}) + \beta_\delta(t_\mathrm{i}^\mathrm{K}, t_\mathrm{f}^\mathrm{K}, t^\mathrm{D}) \times \\
\label{eq:K_COCA}
& & \hfill \quad\quad\quad\quad\quad\quad\quad \left\lbrace \left[ \boldsymbol{\nabla}\left(\Delta^{-1}\delta\right)\right]\!\!(t^\mathrm{D}) - D_1(t^\mathrm{D}) \boldsymbol{\Psi}_1 + \left[ D_2(t^\mathrm{D}) - D_1^2(t^\mathrm{D}) \right] \boldsymbol{\Psi}_2 + \textbf{g}_\mathrm{ML}(t^\mathrm{D}) \right\rbrace . \nonumber
\end{eqnarray}
Using equations \eqref{eq:drift-COCA-standard}--\eqref{eq:kick-COCA-standard}, the standard discretisation of operators \citep{Quinn1997} gives the time prefactors as \citep[][equation (B3)]{Leclercq2020sCOLA},
\begin{equation}
\alpha_\textbf{p}(t_\mathrm{i}^\mathrm{D}, t_\mathrm{f}^\mathrm{D}, t^\mathrm{K}) \equiv \int_{t_\mathrm{i}^\mathrm{D}}^{t_\mathrm{f}^\mathrm{D}} \mathpzc{D}(\tilde{t}) \, {\drm}\tilde{t} = \int_{t_\mathrm{i}^\mathrm{D}}^{t_\mathrm{f}^\mathrm{D}} \frac{{\drm}\tilde{t}}{\tilde{t}^2\mathcal{H}(\tilde{t})}, \quad \beta_\delta(t_\mathrm{i}^\mathrm{K}, t_\mathrm{f}^\mathrm{K}, t^\mathrm{D}) \equiv \int_{t_\mathrm{i}^\mathrm{K}}^{t_\mathrm{f}^\mathrm{K}} \mathpzc{K}(\tilde{t}) \, {\drm}\tilde{t} = -\frac{3}{2} \Omega_\mathrm{m}^{(0)} \mathcal{H}^{(0)2} \int_{t_i^\mathrm{K}}^{t_f^\mathrm{K}} \frac{\drm \tilde{t}}{\tilde{t} \mathcal{H}(\tilde{t})} .
\label{eq:time_prefactors_standard}
\end{equation}
The arbitrary function $u$ of $a$ appearing in equations \eqref{eq:drift-COCA-modified}--\eqref{eq:kick-COCA-modified} can be used to improve upon the standard discretisation of operators \citep[][appendix A]{Tassev2013}. Indeed, if during the time step, the terms between brackets in equations \eqref{eq:drift-COCA-modified}--\eqref{eq:kick-COCA-modified} are closer to constants than the terms between brackets in equations \eqref{eq:drift-COCA-standard}--\eqref{eq:kick-COCA-standard}, the approximation will hold better.
Therefore, using equations \eqref{eq:drift-COCA-modified}--\eqref{eq:kick-COCA-modified}, the modified discretisation of operators gives the time prefactors, for any positive function $u$ of $t$, as \citep[][equation (B11)]{Leclercq2020sCOLA},
\begin{equation}
\alpha_{\textbf{p}}(t_\mathrm{i}^\mathrm{D}, t_\mathrm{f}^\mathrm{D}, t^\mathrm{K}) \equiv \frac{1}{u(t^\mathrm{K})} \int_{t_\mathrm{i}^\mathrm{D}}^{t_\mathrm{f}^\mathrm{D}} \mathpzc{D}(\tilde{t}) u(\tilde{t}) \, \drm \tilde{t}, \quad \beta_{\delta}(t_\mathrm{i}^\mathrm{K}, t_\mathrm{f}^\mathrm{K}, t^\mathrm{D}) \equiv \left[ u(t_\mathrm{f}^\mathrm{K}) - u(t_\mathrm{i}^\mathrm{K}) \right] \times \frac{\mathpzc{K}(t^\mathrm{D})}{\left[\drm u(t)/\drm t \right]\! (t^\mathrm{D})} .
\label{eq:time_prefactors_modified}
\end{equation}
In this paper, consistently with earlier literature, we use $u(t) \equiv a^{n_\mathrm{LPT}}$ with $n_\mathrm{LPT} = -2.5$ \citep{Tassev2013,Leclercq2020sCOLA}.

The ML frame of reference gives particles an acceleration $\textbf{g}_\mathrm{ML}(t^\mathrm{D})$ which should satisfy
\begin{equation}
\int_{t_\mathrm{i}^\mathrm{K}}^{t_\mathrm{f}^\mathrm{K}} \mathpzc{K}(t) \mathpzc{V}[\textbf{x}_\mathrm{ML}](t) \, {\drm}t = \int_{t_\mathrm{i}^\mathrm{K}}^{t_\mathrm{f}^\mathrm{K}} \deriv{u(t)}{t} \times \left\lbrace \frac{\mathpzc{K}(t)}{\drm u(t) / \drm t} \mathpzc{V}[\textbf{x}_\mathrm{ML}](t) \right\rbrace \, {\drm}t \approx \beta_\delta(t_\mathrm{i}^\mathrm{K}, t_\mathrm{f}^\mathrm{K}, t^\mathrm{D}) \textbf{g}_\mathrm{ML}(t^\mathrm{D}) .
\end{equation}

In the standard discretisation, the integral can be approximated by using the value of $\mathpzc{V}[\textbf{x}_\mathrm{ML}](t)$ at $t^\mathrm{D}$ (assuming it is constant during the time step), giving $\beta_\delta(t_\mathrm{i}^\mathrm{K}, t_\mathrm{f}^\mathrm{K}, t^\mathrm{D}) \mathpzc{V}[\textbf{x}_\mathrm{ML}](t^\mathrm{D})$ with the definition of $\beta_\delta(t_\mathrm{i}^\mathrm{K}, t_\mathrm{f}^\mathrm{K}, t^\mathrm{D})$ given in equation \eqref{eq:time_prefactors_standard}. In the modified discretisation, the integral can be approximated using the value of $\frac{\mathpzc{K}(t)}{\drm u(t) / \drm t} \mathpzc{V}[\textbf{x}_\mathrm{ML}](t)$ at $t^\mathrm{D}$, giving also $\beta_\delta(t_\mathrm{i}^\mathrm{K}, t_\mathrm{f}^\mathrm{K}, t^\mathrm{D}) \mathpzc{V}[\textbf{x}_\mathrm{ML}](t^\mathrm{D})$ but with the definition of $\beta_\delta(t_\mathrm{i}^\mathrm{K}, t_\mathrm{f}^\mathrm{K}, t^\mathrm{D})$ given in equation \eqref{eq:time_prefactors_modified}. In both cases, we get
\begin{equation}
\textbf{g}_\mathrm{ML}(t^\mathrm{D}) \equiv \mathpzc{V}[\textbf{x}_\mathrm{ML}](t^\mathrm{D}) .
\end{equation}
But the integral is also:
\begin{equation}
\int_{t_\mathrm{i}^\mathrm{K}}^{t_\mathrm{f}^\mathrm{K}} \mathpzc{K}(t) \mathpzc{V}[\textbf{x}_\mathrm{ML}](t) \, {\drm}t = \int_{t_\mathrm{i}^\mathrm{K}}^{t_\mathrm{f}^\mathrm{K}} - \frac{{\drm}}{{\drm}t} \left( \frac{1}{\mathpzc{D}(t)} \frac{{\drm}\textbf{x}_\mathrm{ML}}{{\drm}t} \right) \, {\drm}t =  \int_{t_\mathrm{i}^\mathrm{K}}^{t_\mathrm{f}^\mathrm{K}} - \frac{\mathrm{d \textbf{p}_\mathrm{ML}}}{{\drm}t} \, {\drm}t = \textbf{p}_\mathrm{ML}(t_\mathrm{i}^\mathrm{K}) - \textbf{p}_\mathrm{ML}(t_\mathrm{f}^\mathrm{K}),
\end{equation}
which gives the alternative form
\begin{equation}
\textbf{g}_\mathrm{ML}(t^\mathrm{D}) \equiv \frac{1}{\beta_\delta(t_\mathrm{i}^\mathrm{K}, t_\mathrm{f}^\mathrm{K}, t^\mathrm{D})} \left[ \textbf{p}_\mathrm{ML}(t_\mathrm{i}^\mathrm{K}) - \textbf{p}_\mathrm{ML}(t_\mathrm{f}^\mathrm{K}) \right].
\label{eq:ML_acceleration}
\end{equation}
As such, to use the COCA Kick and Drift operators (Eqs. \eqref{eq:D_COCA} and \eqref{eq:K_COCA}), one does not require to emulate both $\textbf{p}_\mathrm{ML}$ and $\textbf{g}_\mathrm{ML}$, but one only needs a single emulator (for $\textbf{p}_\mathrm{ML}$), which is evaluated at the kick time steps.

In the end, the time evolution between $t_0$ and $t_{n+1}$ is achieved by applying the following operator to the initial state $\left\lbrace \textbf{x}(t_0),\textbf{p}(t_0) \right\rbrace$:
\begin{equation}
\mathrm{L}_+(t_{n+1}) \mathrm{E}(t_{n+1},t_0) \mathrm{L}_-(t_0),
\end{equation}
where $\mathrm{E}(t_{n+1},t_0)$ is the operator given by (see \cref{fig:leapfrog_sketch})
\begin{equation}
\mathrm{K}(t_{n+1/2},t_{n+1},t_{n+1})\mathrm{D}(t_n,t_{n+1},t_{n+1/2}) \left[ \prod_{i=0}^n \mathrm{K}(t_{i+1/2},t_{i+3/2},t_{i+1}) \mathrm{D}(t_i,t_{i+1},t_{i+1/2}) \right] \mathrm{K}(t_0,t_{1/2},t_0),
\label{eq:operator_E}
\end{equation}
and $L_\pm$ will be defined in \cref{eq:Lpm operator}.

\subsection{Generic Drift and Kick operators for PM, COLA and COCA}

The difference between the COCA Kick and Drift operators and the corresponding COLA operators \citep[][appendices A and B]{Leclercq2020sCOLA} is the last term in each operator. Therefore, we can introduce generic operators, valid for both COLA and COCA: for any external momentum $\textbf{p}_\mathrm{ext}$ and acceleration $\textbf{g}_\mathrm{ext}$,
\begin{eqnarray}
\label{eq:D_COCA_ext}
\mathrm{D}(t_\mathrm{i}^\mathrm{D},t_\mathrm{f}^\mathrm{D},t^\mathrm{K}) : & \quad & \textbf{x}(t_\mathrm{i}^\mathrm{D}) \mapsto \textbf{x}(t_\mathrm{f}^\mathrm{D}) = \textbf{x}(t_\mathrm{i}^\mathrm{D}) + \alpha_\textbf{p}(t_\mathrm{i}^\mathrm{D}, t_\mathrm{f}^\mathrm{D}, t^\mathrm{K}) \textbf{p}_\mathrm{res}\!\left(t^\mathrm{K}\right) + \alpha_\mathrm{LPT1}(t_\mathrm{i}^\mathrm{D}, t_\mathrm{f}^\mathrm{D}, t^\mathrm{K}) \boldsymbol{\Psi}_1 + \alpha_\mathrm{LPT2}(t_\mathrm{i}^\mathrm{D}, t_\mathrm{f}^\mathrm{D}, t^\mathrm{K}) \boldsymbol{\Psi}_2 \nonumber \label{eq:D_generic} \\
& & \hfill \quad\quad\quad\quad\quad\quad\quad + \,\, \alpha_\mathrm{ext}(t_\mathrm{i}^\mathrm{D}, t_\mathrm{f}^\mathrm{D}, t^\mathrm{K}) \textbf{p}_\mathrm{ext}\!\left(t^\mathrm{K}\right), \\
\mathrm{K}(t_\mathrm{i}^\mathrm{K},t_\mathrm{f}^\mathrm{K},t^\mathrm{D}) : & \quad & \textbf{p}_\mathrm{res}(t_\mathrm{i}^\mathrm{K}) \mapsto \textbf{p}_\mathrm{res}(t_\mathrm{f}^\mathrm{K}) = \textbf{p}_\mathrm{res}(t_\mathrm{i}^\mathrm{K}) + \beta_\delta(t_\mathrm{i}^\mathrm{K}, t_\mathrm{f}^\mathrm{K}, t^\mathrm{D}) \textbf{g}_\delta(t^\mathrm{D}) + \beta_\mathrm{LPT1}(t_\mathrm{i}^\mathrm{K}, t_\mathrm{f}^\mathrm{K}, t^\mathrm{D}) \boldsymbol{\Psi}_1 + \beta_\mathrm{LPT2}(t_\mathrm{i}^\mathrm{K}, t_\mathrm{f}^\mathrm{K}, t^\mathrm{D}) \boldsymbol{\Psi}_2 \nonumber \label{eq:K_generic} \\
\label{eq:K_COCA_ext}
& & \hfill \quad\quad\quad\quad\quad\quad\quad + \,\, \beta_\mathrm{ext}(t_\mathrm{i}^\mathrm{K}, t_\mathrm{f}^\mathrm{K}, t^\mathrm{D}) \textbf{g}_\mathrm{ext}(t^\mathrm{D}) ,
\end{eqnarray}
where
\begin{eqnarray}
\alpha_\mathrm{LPT1}(t_\mathrm{i}^\mathrm{D}, t_\mathrm{f}^\mathrm{D}, t^\mathrm{K}) & \equiv & - \left[ D_1 \right]_{t_\mathrm{i}^\mathrm{D}}^{t_\mathrm{f}^\mathrm{D}}, \\
\alpha_\mathrm{LPT2}(t_\mathrm{i}^\mathrm{D}, t_\mathrm{f}^\mathrm{D}, t^\mathrm{K}) & \equiv & \left[ D_2 \right]_{t_\mathrm{i}^\mathrm{D}}^{t_\mathrm{f}^\mathrm{D}}, \\
\alpha_\mathrm{ext}(t_\mathrm{i}^\mathrm{D}, t_\mathrm{f}^\mathrm{D}, t^\mathrm{K}) & \equiv & \alpha_\textbf{p}(t_\mathrm{i}^\mathrm{D}, t_\mathrm{f}^\mathrm{D}, t^\mathrm{K}) \mathrm{~for~COCA~or~} 0 \mathrm{~for~COLA,}\\
\textbf{p}_\mathrm{ext}(t^\mathrm{K}) & = & \textbf{p}_\mathrm{ML}(t^\mathrm{K}) \mathrm{~for~COCA~or~} \textbf{0} \mathrm{~for~COLA,}\\
\textbf{g}_\delta(t^\mathrm{D}) & \equiv & \left[ \boldsymbol{\nabla}\left(\Delta^{-1}\delta\right)\right]\!\!(t^\mathrm{D}),\\
\beta_\mathrm{LPT1}(t_\mathrm{i}^\mathrm{K}, t_\mathrm{f}^\mathrm{K}, t^\mathrm{D}) & = & - \beta_\delta(t_\mathrm{i}^\mathrm{K}, t_\mathrm{f}^\mathrm{K}, t^\mathrm{D}) D_1(t^\mathrm{D}), \\
\beta_\mathrm{LPT2}(t_\mathrm{i}^\mathrm{K}, t_\mathrm{f}^\mathrm{K}, t^\mathrm{D}) & = & \beta_\delta(t_\mathrm{i}^\mathrm{K}, t_\mathrm{f}^\mathrm{K}, t^\mathrm{D}) \left[ D_2(t^\mathrm{D}) - D_1^2(t^\mathrm{D}) \right], \\
\beta_\mathrm{ext}(t_\mathrm{i}^\mathrm{K}, t_\mathrm{f}^\mathrm{K}, t^\mathrm{D}) & = & \beta_\delta(t_\mathrm{i}^\mathrm{K}, t_\mathrm{f}^\mathrm{K}, t^\mathrm{D}) \mathrm{~for~COCA~or~} 0 \mathrm{~for~COLA,}\\
\textbf{g}_\mathrm{ext}(t^\mathrm{D}) & = & \textbf{g}_\mathrm{ML}(t^\mathrm{D}) \mathrm{~for~COCA~or~} \textbf{0} \mathrm{~for~COLA}.
\end{eqnarray}

We note that these operators also remain valid for a standard PM algorithm, by setting $\alpha_\mathrm{LPT1}(t_\mathrm{i}^\mathrm{D}, t_\mathrm{f}^\mathrm{D}, t^\mathrm{K})$, $\alpha_\mathrm{LPT2}(t_\mathrm{i}^\mathrm{D}, t_\mathrm{f}^\mathrm{D}, t^\mathrm{K})$, $\beta_\mathrm{LPT1}(t_\mathrm{i}^\mathrm{K}, t_\mathrm{f}^\mathrm{K}, t^\mathrm{D})$, and $\beta_\mathrm{LPT2}(t_\mathrm{i}^\mathrm{K}, t_\mathrm{f}^\mathrm{K}, t^\mathrm{D})$ to zero.

\subsection{Machine learning prediction of the frame of reference in COCA}

The goal in COCA is to find the frame of reference in which $\textbf{p}_\mathrm{res}$ is as small as possible. Therefore, the machine needs to predict:
\begin{enumerate}
\item At any ``kick'' time step $t^\mathrm{K}$,
\begin{equation}
\label{eq:pML_COCA}
\textbf{p}_\mathrm{ML}\!\left(t^\mathrm{K}\right) \equiv \frac{1}{\alpha_\textbf{p}(t_\mathrm{i}^\mathrm{D}, t_\mathrm{f}^\mathrm{D}, t^\mathrm{K})} \left[ \textbf{x}(t_\mathrm{f}^\mathrm{D})- \textbf{x}(t_\mathrm{i}^\mathrm{D}) - \alpha_\mathrm{LPT1}(t_\mathrm{i}^\mathrm{D}, t_\mathrm{f}^\mathrm{D}, t^\mathrm{K}) \boldsymbol{\Psi}_1 - \alpha_\mathrm{LPT2}(t_\mathrm{i}^\mathrm{D}, t_\mathrm{f}^\mathrm{D}, t^\mathrm{K}) \boldsymbol{\Psi}_2  \right] \equiv \textbf{p}_\mathrm{res}^\mathrm{COLA}\!\left(t^\mathrm{K}\right),
\end{equation}
\noindent that is the momentum residual $\textbf{p}_\mathrm{res}^\mathrm{COLA}\!\left(t^\mathrm{K}\right)$ of COLA \citep[][equation (B5)]{Leclercq2020sCOLA}.
\item At any ``drift'' time step $t^\mathrm{D}$,
\begin{eqnarray}
\textbf{g}_\mathrm{ML}(t^\mathrm{D}) & \equiv & - \frac{1}{\beta_\delta(t_\mathrm{i}^\mathrm{K}, t_\mathrm{f}^\mathrm{K}, t^\mathrm{D})} \left[ \beta_\delta(t_\mathrm{i}^\mathrm{K}, t_\mathrm{f}^\mathrm{K}, t^\mathrm{D}) \textbf{g}_\delta(t^\mathrm{D}) + \beta_\mathrm{LPT1}(t_\mathrm{i}^\mathrm{K}, t_\mathrm{f}^\mathrm{K}, t^\mathrm{D}) \boldsymbol{\Psi}_1 + \beta_\mathrm{LPT2}(t_\mathrm{i}^\mathrm{K}, t_\mathrm{f}^\mathrm{K}, t^\mathrm{D}) \boldsymbol{\Psi}_2 \right] \\
& = & \frac{1}{\beta_\delta(t_\mathrm{i}^\mathrm{K}, t_\mathrm{f}^\mathrm{K}, t^\mathrm{D})} \left[ \textbf{p}_\mathrm{res}^\mathrm{COLA}(t_\mathrm{i}^\mathrm{K}) - \textbf{p}_\mathrm{res}^\mathrm{COLA}(t_\mathrm{f}^\mathrm{K}) \right] \equiv - \textbf{g}_\mathrm{res}^\mathrm{COLA}(t^\mathrm{D}), \label{eq:gML_COCA}
\end{eqnarray}
\noindent that is the residual acceleration $\textbf{g}_\mathrm{res}^\mathrm{COLA}(t^\mathrm{D})$ of COLA \citep[][equation (B6)]{Leclercq2020sCOLA}, up to a minus sign.
\end{enumerate}

From equations \eqref{eq:pML_COCA} and \eqref{eq:gML_COCA}, we see that it is sufficient for the machine to predict the momentum residual $\textbf{p}_\mathrm{res}^\mathrm{COLA}\!\left(t^\mathrm{K}\right)$ of COLA at any ``kick'' time step $t^\mathrm{K}$, as the accelerations $\textbf{g}_\mathrm{res}^\mathrm{COLA}(t^\mathrm{D})$ can be derived from the momenta.

\subsection{Initial and final momenta of particles in COCA}

In the initial conditions, we have $\textbf{p}(t_0) = \textbf{p}_\mathrm{LPT}(t_0) + \textbf{p}_\mathrm{ML}(t_0)$, which means that the momentum residual in the COCA frame of reference, $\textbf{p}_\mathrm{res}(t_0) = \textbf{p}(t_0) - \textbf{p}_\mathrm{LPT}(t_0) - \textbf{p}_\mathrm{ML}(t_0)$, should be initialised to zero. Furthermore, if initial conditions are generated with LPT, the ML contribution is $\textbf{p}_\mathrm{ML}(t_0) = \textbf{0}$ initially and we recover $\textbf{p}_\mathrm{res}(t_0) = \textbf{p}(t_0) - \textbf{p}_\mathrm{LPT}(t_0)$, as in COLA.

At the end, the momentum $\textbf{p}_\mathrm{LPT}(t_{n+1}) + \textbf{p}_\mathrm{ML}(t_{n+1})$ of the COCA frame of reference has to be added to $\textbf{p}_\mathrm{res}(t_{n+1})$ to recover the full momentum of particles, $\textbf{p}(t_{n+1})$. These operations correspond respectively to the $\mathrm{L}_-(t_0) : \textbf{p}(t_0) \mapsto \textbf{p}_\mathrm{res}(t_0)$ and $\mathrm{L}_+(t_{n+1}) : \textbf{p}_\mathrm{res}(t_{n+1}) \mapsto \textbf{p}(t_{n+1})$ operators, given by
\begin{equation}
\label{eq:Lpm operator}
\mathrm{L}_\pm(t) : \quad \textbf{p}(t) \mapsto \textbf{p}(t) \pm \textbf{p}_\mathrm{LPT}(t) \pm \textbf{p}_\mathrm{ML}(t) = \textbf{p}(t) \pm \frac{1}{\mathpzc{D}(t)} \left( -\deriv{D_1}{t} \boldsymbol{\Psi}_1 + \deriv{D_2}{t} \boldsymbol{\Psi}_2 \right) \pm \textbf{p}_\mathrm{ML}(t) .
\end{equation}

\twocolumngrid

\section*{Statement of contribution}

DJB implemented the frame of reference emulator, ran the COCA and COLA simulations, produced the plots, and contributed to the analysis and interpretation of the results and the writing of the paper.
MC contributed to the model equations and the validation of COCA.
LD provided code for the bispectrum analyses, advised on the design of the emulator, and contributed to the interpretation of ML performance results.
FL conceived and designed the study, wrote the model equations, modified the \sbmy{} code to accept any arbitrary frame of reference as input, contributed to the analysis and interpretation of results, edited the manuscript, advised early-career authors, and secured funding.
All authors read and approved the final manuscript.

\section*{Acknowledgements}

We thank
Guilhem Lavaux, Natalia Porqueres, Benjamin Wandelt, and Ewoud Wempe
for useful comments and suggestions.
DJB and LD are supported by the Simons Collaboration on ``Learning the Universe.''
FL and MC acknowledge financial support from the Agence Nationale de la Recherche (ANR) through grant INFOCW, under reference ANR-23-CE46-0006-01. LD acknowledges support by the Swedish Research Council (VR) under the project 2020-05143 – “Deciphering the Dynamics of Cosmic Structure".
This work has received funding from the Centre National d’Etudes Spatiales (CNES).
This work was done within the Aquila Consortium (\url{https://www.aquila-consortium.org/}).

For the purposes of open access, the authors have applied a Creative Commons Attribution (CC BY) licence to any Author Accepted Manuscript version arising.

\section*{References}
\bibliography{references}

\end{document}